\documentclass[useAMS,usenatbib]{mn2e}

\newcommand{\msun}{M$_\odot$}
\newcommand{\rsun}{R$_\odot$}

\newcommand{\ace}{$\alpha$}
\newcommand{\ZT}{\citet{Zorotovic2010}}

\title[The \ace\ formalism]{On the $\alpha$ formalism for the common envelope interaction}

\author[De~Marco et al.]{Orsola De Marco$^{1,2}$\thanks{orsola.demarco@mq.edu.au}, Jean-Claude Passy$^{3,2}$, Maxwell Moe$^4$, Falk Herwig$^3$, \newauthor Mordecai-Mark Mac~Low$^2$, \& Bill Paxton$^5$\\
$^1$Department of Physics \& Astronomy, Macquarie University, Sydney, NSW 2109, Australia\\
$^2$Department of Astrophysics, American Museum of Natural History, New York, NY 10024, USA\\
$^3$Department of Physics and Astronomy, University of Victoria, Victoria, Canada\\
$^4$Department of Astronomy, Harvard University, Cambridge, MA 02138, USA\\
$^5$Kavli Institute for Theoretical Physics, UC Santa Barbara, CA, USA}

\pagerange{\pageref{firstpage}--\pageref{lastpage}} \pubyear{2002}

\def\LaTeX{L\kern-.36em\raise.3ex\hbox{a}\kern-.15em
    T\kern-.1667em\lower.7ex\hbox{E}\kern-.125emX}

\begin{document}
\label{firstpage}

\maketitle

\begin{abstract}

The \ace-formalism is a common way to parametrize the common envelope interaction between a giant star and a more compact companion. The \ace\ parameter describes the fraction of orbital energy released by the companion that is available to eject the giant star's envelope.  By using new, detailed stellar evolutionary calculations we derive a user-friendly prescription for the $\lambda$ parameter and an improved  approximation for the envelope binding energy, thus revising the \ace\ equation. We then determine \ace\ both from simulations and observations in a self consistent manner. By using our own stellar structure models as well as population considerations to reconstruct the primary's parameters at the time of the common envelope interaction, we gain a deeper understanding of the uncertainties. We find that systems with very low values of $q$ (the ratio of the companion's mass to the mass of the primary at the time of the common envelope interaction) have {\it higher} values of \ace.  A fit to the data suggests that lower mass companions are left at comparable or larger orbital separations to more massive companions. We conjecture that lower mass companions take longer than a stellar dynamical time to spiral in to the giant's core, and that this is key to allowing the giant to use its own thermal energy to help unbind its envelope. As a result, although systems with light companions might not have enough orbital energy to unbind the common envelope, they might stimulate a stellar reaction that results in the common envelope ejection.

 \end{abstract}

\begin{keywords}binaries: close ---
                    planetary nebulae: general --- 
                    stars: horizontal branch ---
                    stars: evolution ---
                    stars: white dwarfs
\end{keywords}

\section{Introduction}
\label{sec:introduction}

Common envelope (CE) binary interactions occur when expanding stars transfer mass to a companion at a rate so high that the companion cannot accrete it. This results in the companion being engulfed by the envelope of the primary \citep{Paczynski1976}. The companion's orbital energy and angular momentum are then transferred to the envelope via an as yet poorly characterised mechanism. This can result in the ejection of the envelope and in a much reduced orbital separation. If the companion cannot eject the CE, it merges with the core of the primary. The CE interaction is thought to last for only a few years, and it is therefore likely that we have never witnessed it  directly, although some claims have been made, e.g., for V~Hya \citep{Kahane1996}. The existence of companions in close orbits around evolved stars, whose precursor's radius was larger than today's orbital separation, vouches for such interactions having taken place.

The CE interaction is thus thought to be responsible for short period binaries such as cataclysmic variables \citep[CV;][]{King1988,Warner1995}, close binary central stars of planetary nebula (PN; \citealt{DeMarco2009b}), subdwarf B binaries \citep{MoralesRueda2003,Han2002,Han2003}, low mass X-ray binaries \citep{Charles2006}, the progenitors of Type Ia supernovae \citep{Belczynski2005} and other classes of binaries and single stars thought to have suffered a merger (such as FK~Comae stars; \citealt{Bopp1981}). The specific characteristics exhibited by these binary classes, as well as their relative population sizes are dictated by the period and mass ratio distribution of the progenitor binary population, as well as by  the details of the physical interaction during the CE phase. 

The CE interaction can be parametrised in terms of the binding and orbital energy sources at play \citep[e.g.,][]{Webbink1984,Webbink2008}, and the post-CE period has been expressed as a function of how efficiently orbital energy can be used to unbind the CE. The efficiency parameter, \ace, was thus introduced:

\begin{equation}
\alpha = {E_{\rm bin} \over {\Delta E_{\rm orb}}},
\label{eq:energybalance}
\end{equation}

\noindent where $E_{bin}$ is the gravitational binding energy of the envelope and $\Delta E_{orb}$ is the amount of orbital energy released during the companion's in-spiral.  The expressions used for binding and orbital energies in the literature have varied. As a result, the conclusions reached in the numerous papers discussing the CE interaction by means of the \ace-formalism are difficult to compare. The first motivation of this paper is therefore  to choose a formalism by revisiting past choices. 

Several papers \citep[e.g.,][]{Maxted2006,Afsar2008,Zorotovic2010} use individual post-CE binaries to derive \ace. Their observed primary and secondary masses, together with their orbital periods, provide us with parameters of the post-CE systems. Based on the primary mass, one can reconstruct the mass and radius of the primary at the time of the CE interaction and, with this information, a value of \ace\ can be derived. However, this method has many hidden uncertainties, and the values of \ace\ derived in this way are, once again, only indicative. We therefore use our preferred \ace-formalism to re-evaluate the value of \ace\ for observations and simulations in a homogeneous way and with an increased attention to the sources of uncertainty. 

The values of \ace\ determined by simulations, \citep[e.g.,][]{Sandquist1998} have their own flavour of hidden caveats and are not easily comparable with those derived from observations. We therefore use the better-understood simulations from the literature to gain insight in how the values of \ace\ determined in this way compare with those derived from observations.

In \S~\ref{sec:formalism} we discuss the \ace-formalism in the literature and derive our preferred form. We also discuss the value of $\lambda$, often used in parametrizing the envelope binding energy. In \S~\ref{sec:alpha} we calculate the value of \ace\ for a set of simulations and observations in a self consistent manner. We then (\S~\ref{sec:discussion}) discuss the dependence of \ace\ on stellar and system parameters. We conclude and summarise in \S~\ref{sec:summary}.

\section{The \ace\ equation}
\label{sec:formalism}

%The \ace-formalism is based on the comparison between the envelope binding energy and the orbital energy surrendered by the companion as it spirals deeper and deeper into the primary's potential well (Eq.~\ref{eq:energybalance}). The expressions used for the binding energy of the envelope is however an approximation and as such several forms can be found in the literature. Before using the \ace-formalism we need to determine our preferred expression, which is what we do next.

In this section we determine the best form of the \ace\ equation (Eq.~\ref{eq:energybalance}).

%**************************************************
%  ALPHA FORMALISM IN THE LITERATURE 
%****************************************************

\subsection{The \ace-formalism in the literature}
\label{ssec:literature}

The original \ace\ equation can be found in \citet{Tutukov1979a}, who used:

\begin{equation}
\Psi  {M^2 \over R}= \beta  {{M_{\rm He} m}\over{2 R_{\rm He}}},
\label{eq:tutu}
\end{equation}

\noindent where we have maintained the original symbols to emphasise the subtly different assumptions made in each equation. $\beta$ is the binding energy parameter equivalent to \ace, $M_{\rm He}$  and $R_{\rm He}$ are the post-CE primary's core mass and  radius, respectively, $m$ is the companion's mass, $M$ and $R$ are the mass and radius of the primary at the time of the CE interaction. The value of $\Psi$ was taken to be 0.5 to account for the fact that the radius of the primary at the start of the {\it dynamically-significant} part of the CE interaction was thought to be twice as large as it was at the beginning of the CE interaction. 

Later \citet{Iben1984} used a similar expression but modified the symbols:

\begin{equation}
{M_1^2 \over A}=\alpha {{M_{1,R} M_2}\over{A_f}},
\label{eq:iben}
\end{equation}

\noindent where $M_1$ is the primary mass at the time of the CE and $M_{\rm 1,R}$ is the primary (remnant) mass after the CE interaction. Remembering that there is a factor of 1/2 on both sides of Eq.~\ref{eq:iben} we see that $A$, the initial binary separation, has replaced the initial primary radius (a reasonable assumption) and that  $A_f$, the final binary separation, has taken the place of the primary core radius, a choice that seems sensible, as the final binary separation should be larger than the primary's core radius.  

In the same year,  \citet{Webbink1984} rewrote the expression as:

\begin{equation}
-G\frac{M_1 M_{1,e}}{\lambda R_{1,L} } = -\alpha_{CE} G\left[ \frac{M_{1,c}M_2}{2A_f} - \frac{M_1 M_2}{2A_i}\right] ,
\label{eq:webbink}
\end{equation}

\noindent where we have adopted the symbols used by \citet{Webbink2008}: $M_{1,c}$ and $M_{1,e}$ are the primary's core and envelope masses, respectively, $A_i$ is the inital binary separation, $R_{1,L}$ is the Roche lobe of the primary at the onset of mass transfer and where $\lambda$ is a number of order unity which depends on the mass distribution of the primary's envelope ($\lambda$R is effectively the mass-weighted mean radius of the envelope). 
The symbol $\lambda$ was actually introduced by \citet{deKool1987} as an addition to the \citet{Webbink1984} equation and has been included in the equation ever since. Eq.~\ref{eq:webbink} contains a term for the orbital energy at the beginning of the CE interaction, a term that others have neglected, on the ground that its value is far smaller than that of the final orbital energy, due to the considerable in-spiral that happens during the interaction. 

\citet{Yungelson1993} rewrote the \ace\ equation in the following way:

\begin{equation}
\frac{(M_1-M_{1,R})(M_1+M_2)}{A_0} = -\alpha M_{1,R} M_2 \left[(\frac{1}{A_f} - \frac{1}{A_0})\right] ,
\label{eq:Yun}
\end{equation}

\noindent where all symbols have been previously defined and $A_0$ is the binary separation at the beginning of the {\it dynamically-significant} part of the CE interaction. Eq.~\ref{eq:Yun} actually contains a factor of 0.5 on both sides of the equality, which was cancelled out by \citet{Yungelson1993}, but that must be there for the expression of the orbital energy to be correct. This factor implies that the size of the primary at the start of the dynamically-significant part of the CE interaction is $2 \times A_0$ (Eq.~\ref{eq:iben}). Using this expression will therefore result in smaller values of \ace\ (as also noticed by \citet{Han1995}). Aside from the difference in primary radius, Eq.~\ref{eq:Yun} is different from that of \citet{Webbink1984} also because the binding energy of the envelope is more negative:  the giant's envelope during the CE is bound also by  the gravitational attraction of the companion within it. 

Preemptying our derivation in \S~\ref{ssec:binding}, in this work we will use:
		\begin{equation}
-G\frac{M_e(\frac{M_e}{2} + M_c)}{\lambda R}   = - \alpha G\left[ \frac{M_cM_2}{2A_f} - \frac{(M_c+M_e)M_2}{2A_i}\right]
\label{eq:alphaformalism}
\end{equation}

\noindent where $M_c$ and $M_e$ are now the giant primary's core mass (which is assumed to be the same as the mass of the primary after the CE interaction) and the primary's envelope mass (which is assumed to be ejected by the interaction), respectively. In \S~\ref{ssec:virial} we will use the virial theorem to show that the energy budget should include a factor of 0.5 on the left-hand-side of Eq.~\ref{eq:alphaformalism}, corresponding to the thermal energy of the gas lessening the stellar envelope's gravitational potential well. For now, however, we have presented the \ace\ equation using only the traditionally-included energy sources. The thermal energy source will be discussed further in \S~\ref{ssec:virial}, \S~\ref{ssec:results} and \S~\ref{sec:summary}.

In the case of observed systems (\S~\ref{ssec:obsandalpha}), we assume that the primary is filling its Roche lobe radius at the beginning of the CE interaction (i.e., $R=R_{1,RL}$) and that $A_i=R_{1,RL}/r_{1,RL}$, where \citep{Eggleton1983,Webbink2008}:

\begin{equation}
r_{1,{\rm RL}} \sim {{0.49 q^{-2/3}} \over {0.6 q^{-2/3} + \ln (1 + q^{-1/3})}}
\label{eq:RL}
\end{equation}

\noindent and $q$ = $M_2/M_1$\footnote{Different studies substitute the Roche Lobe radius for $A_i$, instead of  the actual separation between primary and secondary at the time of Roche lobe overflow. The initial orbital energy term is mostly negligible compared to the final orbital energy one though, so specific choices of the initial orbital separation are not fundamental.}.  Our expression is almost identical to that of \citet{Webbink1984}, except that our binding energy is somewhat lower.
 
Clearly, irrespective of which formalism one uses, it is paramount that values of \ace\ calculated with one formalism not be used as input to a different formalism in order to derive, for instance, values for the final orbital separation.

%********************** BINDING ENERGY AND LAMBDA ******************************************

\subsection{The binding energy term}
\label{ssec:binding}

Several authors \citep[e.g.,][]{Han1995} chose not to use an approximation for the binding energy of the envelope, but instead integrate the envelope mass from stellar structure calculations. As we will see in \S~\ref{ssec:lambda} this choice does not necessarily lead to a more accurate value of \ace. So, here we determine as accurate an approximation to the binding energy as possible.

The binding energy of a star is:

\begin{equation}
	E_{\rm bin} \equiv - \int \frac{Gm}{r}dm .
\label{eq:defbind}
\end{equation}

\noindent In the case of giants, the small and dense core, is surrounded by a vast envelope whose density rapidly falls with radius. Therefore we model the core as a point mass and the envelope as a shell of homogeneous density: 

$$ \rho = \frac{M_e}{4\pi (\lambda R)^2 }, $$

\noindent located at a distance $\lambda R$ from the core. The factor $\lambda < 1$ accounts for the fact that, in the shell approximation of the real envelope, the shell is located at the mass-averaged stellar radius. Using this assumption we calculate separately the envelope binding energy deriving from the attraction of the core, $E_{\rm ce}$, and that deriving from the attraction of the envelope onto itself, $E_{\rm ee}$:

\begin{equation} 
E_{\rm bin} = E_{\rm ce} +  E_{\rm ee}.
\label{eq:BinEnSplit}
\end{equation}

\noindent The core-envelope binding energy can be easily calculated:

\begin{equation}
	E_{\rm ce} = -G \int\frac{M_c dm}{\lambda R} = -G\frac{M_cM_e}{\lambda R},
\end{equation}

\noindent where $dm$ is a parcel of envelope mass sitting at distance $\lambda R$ from the core, and the integration is over the entire envelope, approximated as a shell with radius $\lambda R$. The self-gravity of the envelope results in:

\begin{equation}
	E_{\rm ee} = - \frac{G}{2}\int\int\frac{dm^{'} dm}{|\mathbf{r} - \mathbf{r^{'}}|} ,
\label{eq:BinEnEE}	
\end{equation}

\noindent where $dm$ and $dm^{'}$ are mass parcels in the envelope shell separated by a distance $|\mathbf{r} - \mathbf{r^{'}}|$. The factor $\frac{1}{2}$ prevents double counting the interaction between parcels. Since we have made the assumption that all mass parcels are in a shell with radius $\lambda R$,   $|\mathbf{r^{'}}| = |\mathbf{r}| = \lambda R$, the denominator is:

\begin{equation}
\frac{1}{|\mathbf{r} - \mathbf{r^{'}}|} = \frac{1}{\lambda R} \sum_{n=0}^{\infty} P_n(\cos{\psi}) \sim  \frac{1}{\lambda R} \left(1 - (\cos{\psi})\right),
\label{eq:BinEnAnotherEq}
\end{equation}
 
\noindent where $P_n$ is the Legendre polynomial. For the final step, we only consider $P_0 = 1$ and  $P_1 = \cos{\psi}$, where $\psi$ is the angle between ${\bf r}$ and ${\bf r^{'}}$ and we neglect orders $n>1$. Because of symmetry, the term including $\cos \psi$ vanishes. Substituting Eq.~\ref{eq:BinEnAnotherEq} into Eq.~\ref{eq:BinEnEE}, we obtain the binding energy of the envelope in the shell approximation:

%$$ \int\int \frac{1}{R}dmdm^{'} = \frac{M_e^2}{R} \qquad , \qquad \int\int \frac{\cos{\psi}}{R}dmdm^{'} = 0 $$

%because the origin of the reference frame is the mass center.

\begin{equation}
	E_{\rm bin} \sim - G\frac{M_e(\frac{M_e}{2} + M_c)}{\lambda R}.
\label{eq:binlambda}
\end{equation}

This expression is different from all of the ones used before (the numerator of our expression is the lowest) although our expression is actually quite similar to that used by \citet{Webbink1984}.  
%The next question is what the value of $\lambda$ should be and how much it is a function of a given stellar structure.  

% ***********************************  VIRIAL ******************************************

\subsection{The thermal energy}
\label{ssec:virial}

The virial theorem can be used to quantify  another source of energy that may play a role in the CE interaction, namely, the thermal energy of the envelope \citep{Webbink2008}. The familiar identity $2K + U$=0, where $K$ is the total thermal, or kinetic energy\footnote{Throughout this paper we will keep referring to the energy due to the thermal motion of the particles as the thermal energy. This will avoid confusion with the energy stored in the bonds of atoms and molecules, called the internal energy by \citet{Webbink2008}.} of the star and $U$ is its potential (binding) energy, accurately represent the global properties of the entire star.  If we only include the stellar envelope, rather than the entire star, the virial theorem takes a slightly different form (for the derivation see \citet{Webbink2008}):

\begin{equation}
2K_{\rm env} + 4 \pi R_c^3 P_c + E_{\rm bin,env} = 0,
\label{eq:internal-envelope}
\end{equation}

\noindent where $K_{\rm env}$ is the thermal energy of the envelope only, and we now use $E_{\rm bin,env}$ to represent the binding energy of the envelope only; $R_c$ and $P_c$ are the values of the radius and the pressure at the core-envelope boundary, respectively (defined in \S~\ref{ssec:lambda}).  

We can show that the extra term ($4 \pi R_c^3 P_c$) can be neglected, by integrating stellar structure models (whose details will be given in  \S~\ref{ssec:lambda}). We use a 2-\msun\ main sequence model evolved to the RGB and AGB phases. The total stellar binding energy is integrated using the denominator in Eq.~\ref{eq:integration}. A similar equation is used to determine the total thermal energy of the stellar models:

\begin{equation}
U = \Sigma^R_{r_i = R_c}  { 3 \over 2} P(r_i)4 \pi r_i^2 \Delta r_i,
\label{eq:internal}
\end{equation}

\noindent where $r_i$ is the radius of the $i^{th}$ concentric shell, $\Delta r_i$ are the shell's thickness, and $P(r_i)$ is the pressure in the shell. $R_c$ is the core-envelope boundary radius chosen according to the criteria explained in  \S~\ref{ssec:lambda}.

\begin{table}
\begin{center}
	{\begin{tabular}{ccc}
	\hline
	  & RGB & AGB\\
	\hline
	 $2U_{\rm env}$ (ergs) & $0.9983 \times 10^{48}$ &  $2.7101 \times 10^{47}$ \\
	 $4 \pi R_c^3 P_c$ (ergs) & $0.0532 \times 10^{48}$ &  $ 0.1365 \times 10^{47}$ \\
	 $ 2U_{\rm env} + 4 \pi R_c^3 P_c$ (ergs) & $1.0516 \times 10^{48}$ &  $ 2.8466 \times 10^{47}$ \\
	 $E_{\rm bin,env}$ (ergs) & $-1.0433 \times  10^{48}$ & $ -2.8348  \times 10^{47}$  \\
	 error without extra term  & 4 \% & 4 \%   \\ 
	 error with extra term & 0.8 \% & 0.4 \%   \\
	\hline
	\end{tabular}}
	\caption{Error on the virial theorem calculations for the entire star using the 2 \msun \ model. $R_c$ defined by the left over mass criterion in \S~\ref{ssec:lambda}.}
	\label{tab:internal-envelope}
\end{center}
\end{table}

We see  (Table~\ref{tab:internal-envelope}) that the first two terms in Eq.~\ref{eq:internal-envelope} do equate to the third term to within less than 1\% for both the RGB and AGB models. We also see that not including the second term, still results in an acceptable equality between twice the thermal energy and the gravitational binding energy of the envelope. In this way we have a very simple way to account for the thermal energy of the stellar envelope in the CE energy budget: the thermal energy is simply one half of the binding energy. Since the two energy sources have opposite signs, the effect of accounting for the thermal energy is to ``fill" the stellar envelope potential well, or make the envelope lighter. 

Finally, one should remember that there is another possible energy source: the dissociation and ionisation energy of the envelope. This was discussed by \citet{Han1995} and \citet{Webbink2008}. We will return to this topic in \S~\ref{sec:discussion}.

%**********************  LAMBDA ******************************************

\subsection{The stellar structure parameter $\lambda$}
\label{ssec:lambda}

%One may argue that it is not important to know the value of $\lambda$ because one should simply integrate the binding energy of the envelope using the stellar structure calculation pertinent to the particular CE interaction under study. This approach, however, is not practical because specific stellar structure calculations for a primary of interest are not always available. It is also arguable whether it would result in higher accuracy. First of all we never actually know the exact parameters of the primary at the time of the CE interaction. Second, the binding energy of the envelope from direct integration of the stellar structure depends sensitively on the adopted core radius, and the location of the core-envelope boundary is itself the subject of debate. 

To determine a suitable value of $\lambda$ we use a 2-\msun\ model calculated with the code Modules for Experiments in Stellar Astrophysics (MESA\footnote{mesa.sourceforge.net}; \citealt{Paxton2010}), evolving from the main sequence to the tip of the AGB.  %MESA is a new, modular, fast stellar evolution code that solves the mixing, burning and structure operators simultaneously. A network including hydrogen- and helium-burning has been used, as is appropriate for the evolution of low- and intermediate mass stars. A detailed comparison of MESA code models with AGB stellar evolution models from the EVOL code
This new stellar evolutionary code compares well with EVOL \citep[e.g.][]{Herwig2004}.

By equating the envelope binding energy obtained by numerical integration of the stellar structure and that obtained by the approximation in Eq.~\ref{eq:binlambda}, we calculate $\lambda$ in this way:

\begin{equation}
\lambda = {{- (G M_e / R) (M_e / 2 + M_c)}\over{-4\pi G \sum_{r_i=R_c}^{R} m_{\rm int}(r_i)\rho_i r_i\Delta r_i}},
\label{eq:integration}
\end{equation}

\noindent where $\rho_i$ is the density and radius of a shell of material of thickness $\Delta r_i$ and $m_{\rm int}$ is the mass internal to that radius.

%The radius at which the core ends and the envelope starts must be chosen carefully, since the density of envelope matter near the core is high, so small changes in radius correspond to large changes of the total binding energy and hence in the value of $\lambda$.  
The choice of a core radius based only on density (e.g., where $dm/d\log r$ reaches a minimum, see also \citealt{Tauris2001} and \citealt{Bisscheroux1998}) locates a core-envelope boundary that is quite different from any boundary located on physical grounds. We therefore chose
the core-envelope boundary to be at the 
%There are three criteria that one could adopt to determine the location of the core boundary on physical grounds: (i) the radius where the abundance of hydrogen is 0.1 (this is the criterion adopted by \citet{Dewi2000}), (ii) the largest radius where the nuclear reaction rate decreases below a given threshold value  (just outside the hydrogen-burning shell) or (iii) 
radius where the nuclear burning reaches a maximum {\it plus} the thickness of a remaining envelope of mass $10^{-3}$~\msun\ for the 2.0-\msun\ AGB model described above and $10^{-2}$~\msun\ for the corresponding RGB model. This criterion is motivated by the fact that AGB and RGB stars depart their respective giant branches by contracting, when the envelope mass decreases below a threshold value \citep[or, for RGB stars, when core helium is ignited;][]{Bloecker1995,Castellani2006}. In a CE situation, such contraction would detach the primary from its Roche lobe and dictate the end of the interaction. This criterion leads to very similar core-envelope boundaries and values of  $\lambda$ to the criteria of \citet{Dewi2000}. For more discussion see Appendix~\ref{appendixA}.

%In Table~\ref{tab:lambda} we list the values of the core mass, core radius and hydrogen abundance of the 2-\msun\ model mentioned above for two key evolutionary phases, when the star is half way up the RGB (M=2.0~\msun, R=20~\rsun\ and $\tau$=1.0~Gyr) and at the start of the thermally-pulsating AGB (M=1.93~\msun, R=100~\rsun\ and $\tau$=1.2 Gyr), along with the derived values of the envelope binding energy and of $\lambda$. As we can see (Fig~\ref{fig:CoreBoundary} right-most vertical line in all panels), the density-based criterion leads to core radius values which are well outside any of locations in the star that could plausibly be defined as the core-envelope boundary. For this reason we abandon this criterion. The hydrogen abundance criterion leads to a core radius in the middle of the hydrogen burning shell, while the other two criteria lead to slightly larger, and similar, radii. Our preferred criterion is the ``remnant envelope" criterion because it makes the most sense in a CE situation where the CE is terminated by the stellar contraction, which happens when the envelope mass left is the same as the ``remnant envelope" mentioned above.

With this choice of core-envelope boundary, we carried out stellar evolution calculations for main sequence masses of 1, 1.5, 2, 2.5, 3, 4 and 5~\msun. For each of these models, we determined $\lambda$ for a few stellar structure models clustered on the RGB and AGB. The results are summarised in Table~2. For each main sequence mass, we list the numbers of stellar structures used in determining the average value of $\lambda$, the smallest and largest stellar radius for the models in each cluster and finally the average value of lambda as well as the 1-$\sigma$ spread of values. The minimum and maximum radii on the RGB and AGB for each model cluster were chosen as follows: the minimum radii were those of the first model that had developed a clear shell burning structure (hydrogen for the RGB and helium for the AGB). The maximum values were the maximum RGB and AGB radius values for the respective models. The value of $\lambda$ does not vary significantly within each cluster, despite the relatively large radius range, because the mass is centrally concentrated. 

\begin{figure}
\vspace{7cm}
	\begin{center}
	\includegraphics{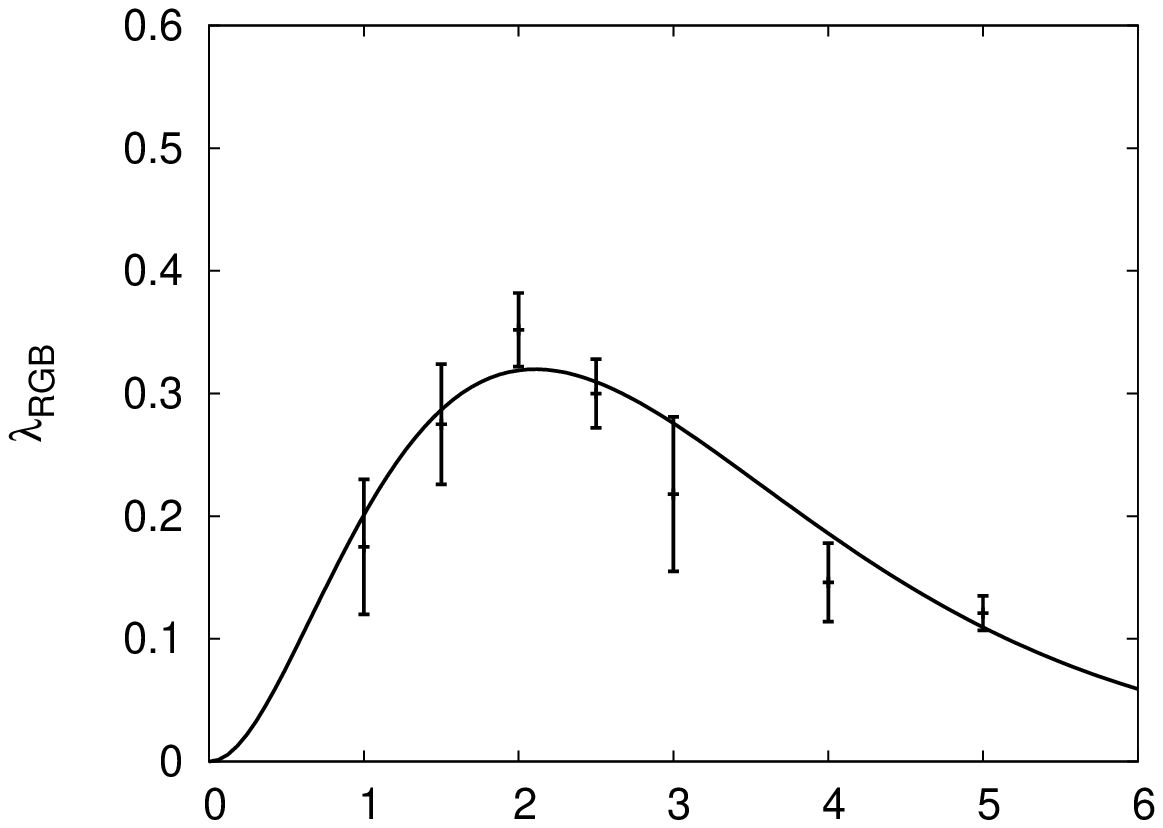}
	\includegraphics{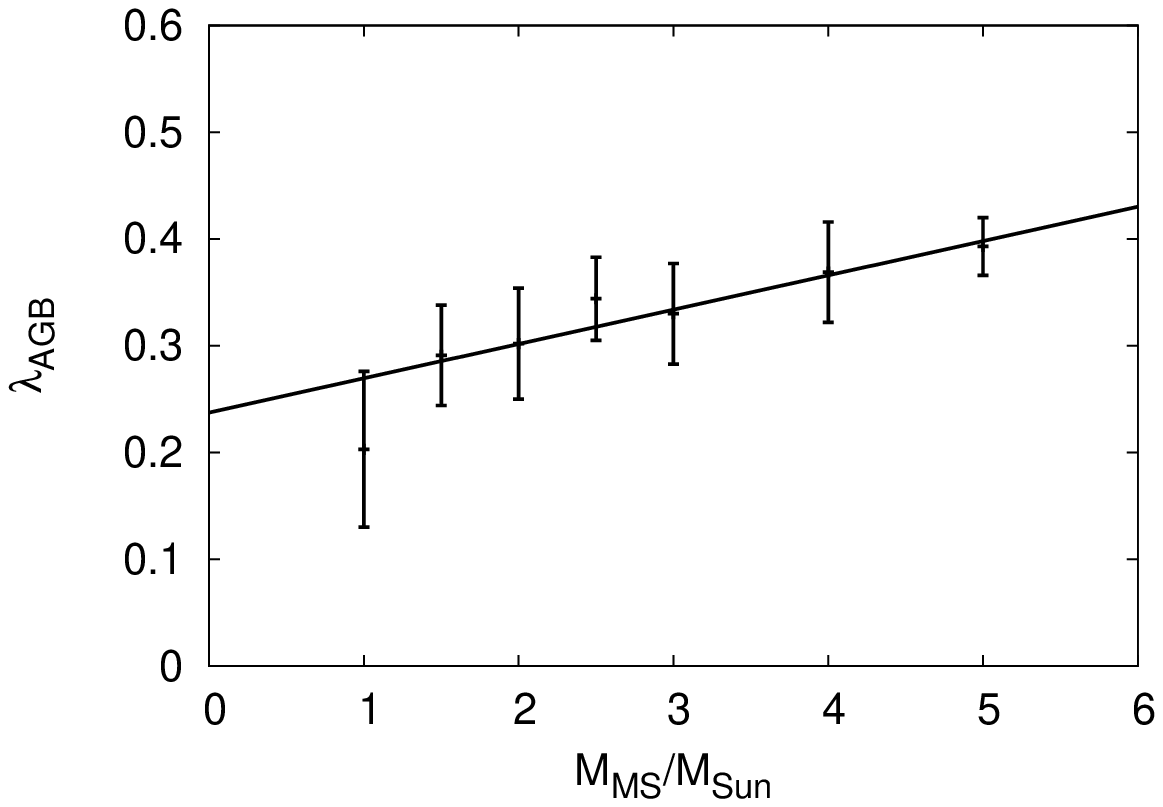}
	\end{center}\caption{The average value of $\lambda$ (symbols with error bars) calculated for clusters of models on the RGB (top panel) and AGB (bottom panel) as a function of main sequence mass, along with the best fits (solid lines)}
	\label{fig:lambdafits}
\end{figure}

We have fitted the average $\lambda$ values as a function of main sequence mass (Fig.~\ref{fig:lambdafits}) and obtained the resulting fit parameters:

\begin{equation}
\lambda_{\rm RGB} = (0.547 \pm 0.068) \times \left({M_{\rm MS} \over {\rm M}_\odot}\right)^{(2.11 \pm 0.12)}
\times \exp{(-M_{\rm MS}/M_\odot)}
\label{eq:lambdaRGB}
\end{equation}

\noindent and 

\begin{equation}
\lambda_{\rm AGB} = (0.237 \pm 0.021) + (0.032 \pm 0.006) \times {M_{\rm MS} \over {\rm M}_\odot},
\label{eq:lambdaAGB}
\end{equation}

\noindent which are valid for 1.0~\msun $\le M_{\rm MS} \le 5.0$~\msun.  By calculating the value of $\lambda$ for each of our systems using the expression:

\[
\lambda^{-1} \sim 3.000 - 3.816 M_e + 1.041 M_e^2 + 0.067 M_e^3 + 0.136 M_e^4
\]

\noindent of \citet[][]{Webbink2008}, we obtain much larger values.

\begin{table}
\begin{center}
	{\begin{tabular}{ccccccc}
	\hline
	$M_{\rm MS}$ & \# of models & $R_{\rm min}$ & $R_{\rm max}$  & $\lambda$ \\
         (\msun) & in cluster  & (\rsun) & (\rsun)  &   \\

	\hline
		\multicolumn{5}{c}{RGB} \\

	\hline
	1   & 6 & 20   & 120   & $0.175 \pm 0.055$ \\
 	1.5   & 6 & 20   & 120   &  $0.275 \pm 0.049$ \\
 	2   & 4 & 20   & 80   &  $0.352 \pm 0.030$ \\
 	2.5   & 5 & 10   & 30   &  $0.300 \pm 0.028$ \\
 	3   & 4 & 10   & 40   &  $0.218 \pm 0.063$ \\
 	4   & 5 & 10   & 50   &  $0.146 \pm 0.032$ \\
 	5   & 4 & 20   & 80   &  $0.121 \pm 0.014$ \\
 	\hline
	\multicolumn{5}{c}{AGB} \\
	\hline
	1   & 4 & 150   & 300    &  $0.203 \pm 0.073$ \\
 	1.5   & 6 & 150   & 400   & $0.291 \pm 0.047$ \\
 	2   & 7 & 50   & 350   &  $0.302 \pm 0.052$ \\
 	2.5   & 6 & 50   & 300   &  $0.344 \pm 0.039$ \\
 	3   & 7 & 50   & 350   &  $0.330 \pm 0.047$ \\
 	4   & 9 & 100   & 500   & $0.369 \pm 0.047$ \\
 	5   & 8 & 150   & 500   &  $0.393 \pm 0.027$ \\
	\hline 
	\end{tabular}}
	\label{tab:lambdamodels}
	\caption{Average values of $\lambda$ calculated for clusters of models equally distributed in radius between $R_{\rm min}$ and $R_{\rm max}$, on the RGB and AGB}
\end{center}
\end{table}

Finally, we comment on the fact that in our static calculation of the binding energy we assume that the envelope that needs to be ejected is the same as the mass above the core-envelope boundary just discussed. This assumption does not take into account that in an evolving, mass-losing and expanding giant, the actual mass left on the star after termination of the AGB may not be exactly that which we accounted for in our static calculation. \citet{Deloye2010} calculate the adiabatic mass-loss response of the primary. They explain that massive stars - they show the case of a 10 \msun\ - have a shallow entropy gradient in the core/envelope region, and this can lead to a 30\% difference between the donor's post-CE remnant mass and $M_c$. However, for the lower mass stars we are interested in, the gradient is much steeper and two masses are similar. The other concern is that the binding energy as we calculate it, does not take into account the work done by the envelope throughout the adiabatic mass-loss. However, according to \citet{Ge2010}, including this effect does not make a large difference on the binding energy (see their Section 4 and their equations 62 and 83).
%Whether they calculate the binding energy using eq (62) or eq (83):
%" In practice, we Þnd that the two estimates are in close agreement at the transition, and use a weighted mean of the two to determine best estimates of ÆE1 and E1f ."

% =============================================================================
%							ALPHA
% =============================================================================

\section{The determination of \ace\ using simulations and observations}
\label{sec:alpha}

Here we calculate \ace\ for observed post-CE binary systems as well as for a set of CE simulations. To do so, we use our approximation (Eq.~\ref{eq:alphaformalism}) together with the value of $\lambda$ determined from the fits in \S~\ref{ssec:lambda} (Eqs.~\ref{eq:lambdaRGB} and \ref{eq:lambdaAGB}).

% =============================================================================
%							OBS AND ALPHA
% =============================================================================

\subsection{The pre-CE giant reconstruction technique}%\\ Obtaining $M_1$ and $R_i$ from $M_c$}
\label{ssec:reconstruction}

%The determination of \ace\ from known post-CE binary systems has advantages and disadvantages. The post-CE separation and the masses of the components can be reasonably well determined from observations, but the initial conditions ($M_1$ and $R$) have to be derived using  theoretical knowledge of stellar structure and evolution.   

We assume that observed post-CE systems with $M_c \ge 0.47$~\msun\ have suffered a CE interaction on the AGB and those with $M_c < 0.47$~\msun\ have suffered an interaction on the RGB. The latter assumption is not strictly correct, because more massive primaries can have a  core mass $>$0.47~\msun\ during the RGB ascent (see for instance the 5-\msun\ model in Fig.~\ref{fig:tracks}). However, our assumption is approximately correct because lower mass stars ($\la 2.5$~\msun), which are statistically more common, have cores in the 0.47-0.62~\msun\ range only during their AGB ascent. In \S~\ref{sssec:anti-correlation}  we discuss this topic further.

%These primaries are either helium WDs or have ignited helium after the CE thus becoming CO WDs \citep{Iben1986} {\bf I am not actually sure about this statement any more}. 
Of the post-AGB primaries ($M_c \ge 0.47$~\msun), some are surrounded by a PN, guaranteeing that the system only emerged from the CE interaction recently (or the PN, which has a lifetime of $\la$100\,000 years, would not be visible). When a PN is lacking, there is a chance that the time since the CE interaction might be sufficiently long to have allowed the evolution of the system's orbital period.  In order to avoid such systems we use the binary post-CE list of \citet{Schreiber2003} and \citet{Zorotovic2010}, who determined which post-CE systems have the same period today as they had when they emerged from the CE.  

Once the RGB or AGB origin of the primary is determined, the radius and mass of the giant at the time of the CE interaction can be estimated. For post-RGB primaries, the radius has traditionally been obtained by the core mass-radius relation of \citet[][$R \sim 10^{3.5} M_c^4$]{Iben1985}. RGB masses at the time of the CE interaction cannot be determined from the core mass alone, so a range has to be adopted  \citep[e.g.,][]{Nelemans2000} 

To determine the pre-CE radius and mass for primaries that went through the interaction on the AGB, several studies, \citep[e.g.,][]{Afsar2008} used the core mass-luminosity relation for post-AGB stars \citep[e.g.,][$L/$L$_\odot=56694(M_c$/M$_\odot-0.5$)]{Vassiliadis1994} to determine the luminosity of the primary at the time of the CE interaction. They then used the initial-to-final mass relation to determine the main sequence progenitor mass of today's primary, to then determine the mass of the primary at the time of the CE interaction \citep[e.g.,][]{Vassiliadis1994}. Once the primary's mass and luminosity at the time of the CE interaction have been determined, the radius can be found using the fitting relation of \citet[][$R=1.125 M_1^{-0.33}(L^{0.4}+0.383L^{0.76})$]{Hurley2000}.  This method cannot be applied to primary stars with core mass $\la 0.55$~\msun, because below that value the core mass-luminosity relation of \citet{Vassiliadis1994} becomes extremely imprecise and for core masses $\le 0.5$~\msun, the stellar luminosity becomes negative.  Finally, using the initial-to-final mass relation in this way is incorrect, because the CE interaction interrupted the evolution of the primary and the growth of its core, resulting in a less massive core (i.e., smaller final mass) than if the star had been single.

By studying the relations used in the past one realises that reconstructing the giant's mass and radius using only the mass of the post-CE primary (assumed to have been the mass of the core of the giant), is fundamentally a statistical process. Each relation appears to state that the error on a given quantity, e.g., the giant radius at the time of the CE interaction, depends solely on the error on the measured core mass. However, this error propagation does not include the error in the fit, i.e., the error in the relations themselves. Nor is the error in the assumptions used to establish the relations quantified. We therefore diverge from previous methods, and recalculate the fitting relations from stellar evolution models. This allows us to better quantify the true uncertainty on the entire reconstruction process. In our reconstruction technique, we make use of two sets of stellar evolution calculations. The first set is from the detailed calculations of MESA for $Z=0.01$ (\S~\ref{ssec:lambda}). The second set is from the computations of \citet{Bertelli2008}, who calculated an extensive grid of models including their metallicity dependence.

\subsubsection{The primary mass at the time of the CE interaction}
\label{sssec:mass}

\begin{figure}
\vspace{4.5cm}
	\begin{center}
	\includegraphics{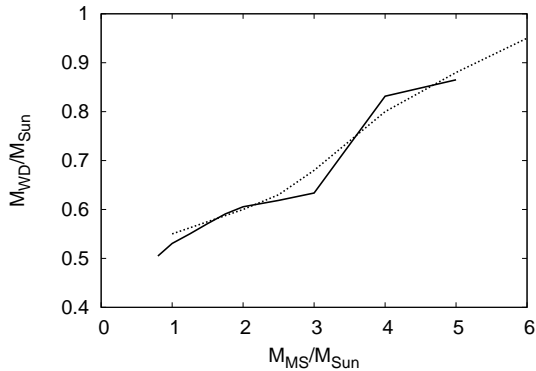}
 	\end{center}\caption{A comparison of the MESA code initial-to-final mass relation (solid line) and that determined empirically by \citet[][dashed line]{Weidemann2000}.}
	\label{fig:MESA-IFMR}
\end{figure}
\begin{figure*}
\vspace{5.5cm}
	\begin{center}
	\includegraphics{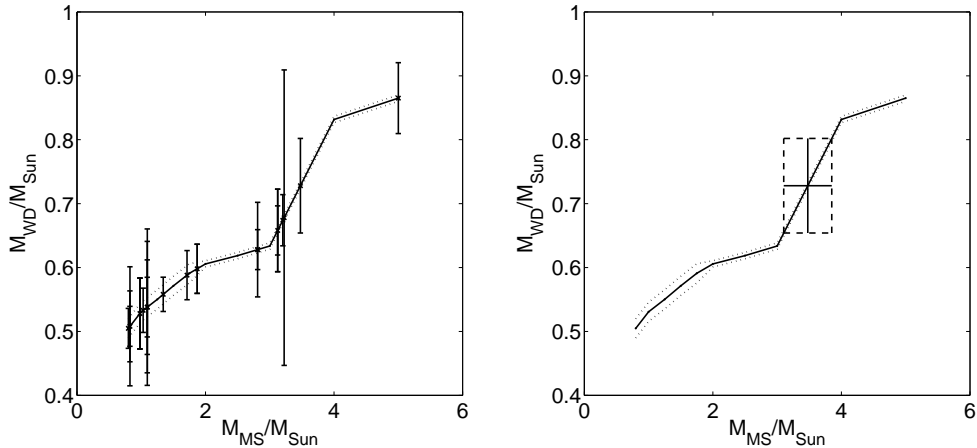}
  	\end{center}\caption{Left: the initial-to-final mass relation (IFMR) determined from the MESA code calculations, plotted together with the data and their error bars (Table~\ref{tab:alphainputs}). The dotted lines delimit the estimated error on the IFMR itself. Right: the IFMR plotted with the data point for BE~UMa to exemplify our determination of the error on the derived main sequence masses.}
	\label{fig:IFMR}
\end{figure*}
To determine the mass of the primary at the time of the CE interaction for the post-AGB group, we start by determining the mass of the main sequence progenitors. To do so we use the initial-to-final mass relation (IFMR) determined with the MESA code. The final mass is assumed to be the mass of the core at the first thermal pulse. Although this may appear as an under estimate of the final mass, it is accurate for core mass values $\la0.52$~\msun, because low mass stars with low mass cores have low mass envelopes which get depleted rapidly by mass loss and do not give the core time to grow. Our assumption also results in good accuracy for masses $\ga 0.58$~\msun, because high mass stars suffer substantial dredge-up so that the core does not grow appreciably during thermal pulses. For values between 0.52 and 0.58~\msun, the core could grow slightly above the value that we are using, resulting in a slight overestimate of the initial mass. However this uncertainty is very small compared  to the very large uncertainty on the core mass, as we explain below.

Older predictions for the IFMR \citep[e.g.,][]{Bloecker1995,Vassiliadis1994} compare well to the MESA IFMR for lower initial masses ($\la$3~\msun), while for the larger initial masses the more recent MESA IFMR yields higher WD masses (also observed by \citet{Han1994}). The reason is that the new MESA models include convective core overshooting during the main sequence, which increases the core mass predictions compared to the older models. However, at low initial mass, models now show efficient third dredge-up, which compensates for the core mass increase from core overshooting. For more massive AGB stars the third dredge-up does not play a significant role in the core mass prediction, and the core mass difference from core overshooting assumptions can be observed. The MESA IFMR compares well with the observationally derived IFMR of \citet[][Fig.~\ref{fig:MESA-IFMR}]{Weidemann2000}, although the observational IFMR has a large scatter. %The resolution of the \citet{Weidemann2000} relation is low, while that of our MESA IMFR relation is higher. Because of the consistency of the two, we decide to use the MESA-calculated relation (for further discussion on the determination of the IFMR see \citet{Weidemann2000}).  

For each system's primary we   obtained the main sequence mass of its progenitor by direct interpolation. The metallicity dependence of the IFMR is small, as can be seen by a fit to the \citet[][]{Bertelli2008} models ($M_{WD}/M_{\odot} = 0.451 + 0.091 M_{\rm MS}/ M_{\odot} - 0.044 \log Z/Z_{\odot}$) and we ignore it.
%by fitting  the stellar evolutionary models of \citet{Bertelli2008} in the mass range 0.8--2.5~\msun\ and for the metallicity range, $\log Z/Z_\odot = -0.9$ to $0.4$\footnote{$M_{WD}/M_{\odot} = 0.451 + 0.091 M_{MS}/ M_{\odot} - 0.044 \log Z/Z_{\odot}$ -- where $M_{WD}$ and $M_{MS}$ are the final, WD mass and main sequence mass, respectively. This relation is very similar to that of \citet{Vassiliadis1994}.}. 
Before we can use the IFMR, we need to account for the fact that the IFMR relates the final WD mass, $M_{\rm WD}$, to its main sequence mass, $M_{\rm MS}$, only if the star evolved to the {\it natural} termination of its AGB life. The CE interaction interrupted the regular AGB evolution and core growth, so that the post-CE primary mass, $M_c$, is smaller than it would have been had the primary been single ($M_{\rm WD}$). This mass discrepancy is determined for each \citet{Bertelli2008} model, assuming that a CE will terminate the AGB evolution at a random value of $R$, between the maximum RGB radius and the maximum AGB radius. We use \citet{Bertelli2008} models in the mass range 0.8--2.5~\msun,  and metallicity range $Z=10^{-4}-0.07$. On average, models depart the AGB with a mass that is 0.028$\pm$0.024~\msun\ lower than if the AGB evolution had progressed till its natural termination. The error in this value is dominated by the lack of knowledge of the precise stage at which the system departs the AGB.  

We can now calculate the WD mass of each of our post-AGB, post-CE primaries and use the IFMR to determine the main sequence mass of its progenitor. The error on the WD masses is determined from the error on the measured post-CE primary masses and the error on the core growth estimation, added in quadrature. In Fig.~\ref{fig:IFMR} (left) we show the MESA IFMR with our interpolated data points and their error bars. To determine the error on the main sequence masses thus obtained, we projected the errors calculated on $M_{\rm WD}$ on the IFMR itself, as demonstrated in Fig.~\ref{fig:IFMR} (right). For three data points at the higher (V471~Tau) and lower (SDSS~J0110+1326 and RR~Cae) extremes of the IFMR, the errors were assumed to be the same as the {\it relative} error of HS~0705+6700 and SDSS~J1548+4057, for the low and high mass, respectively, because these two systems have the same measurement errors and similar core mass values. 
%we projected the WD mass upper error bars to the right, and the lower error bars to the left and determined the value of the main sequence masses of the intercepts, thus obtaining an upper and lower main sequence mass limits for each data point. We then averaged the lower and upper error bars on the main sequence mass obtained in this way. 
Using this error estimation method, we are not taking into account the error on the IFMR itself (plotted as a dotted curve in Fig.~\ref{fig:IFMR},  from approximate estimates). If we had, the errors on the determined main sequence masses would have been only marginally larger.  

\begin{figure*}
\vspace{6cm}
	\begin{center}
	\includegraphics{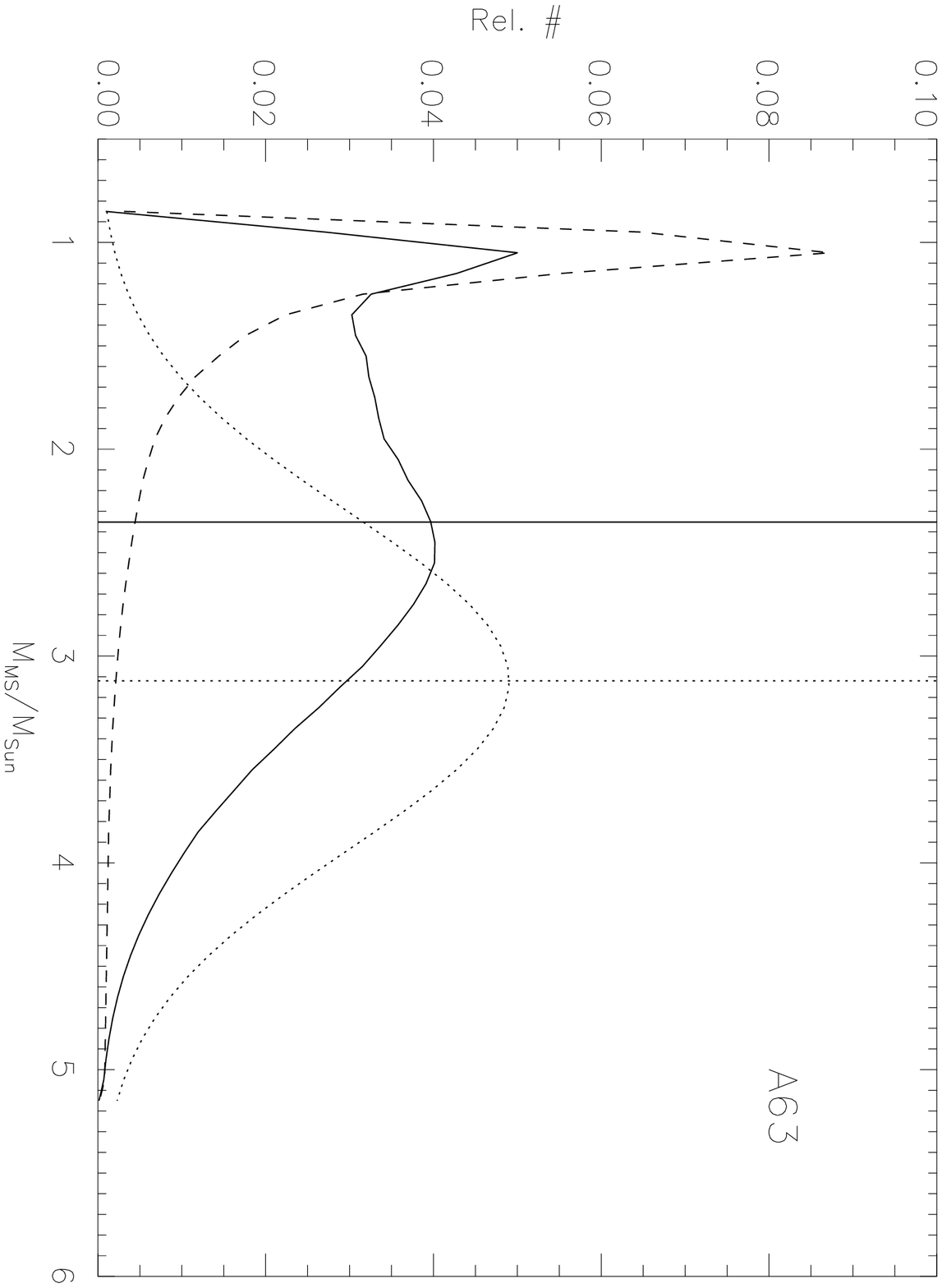}
	\includegraphics{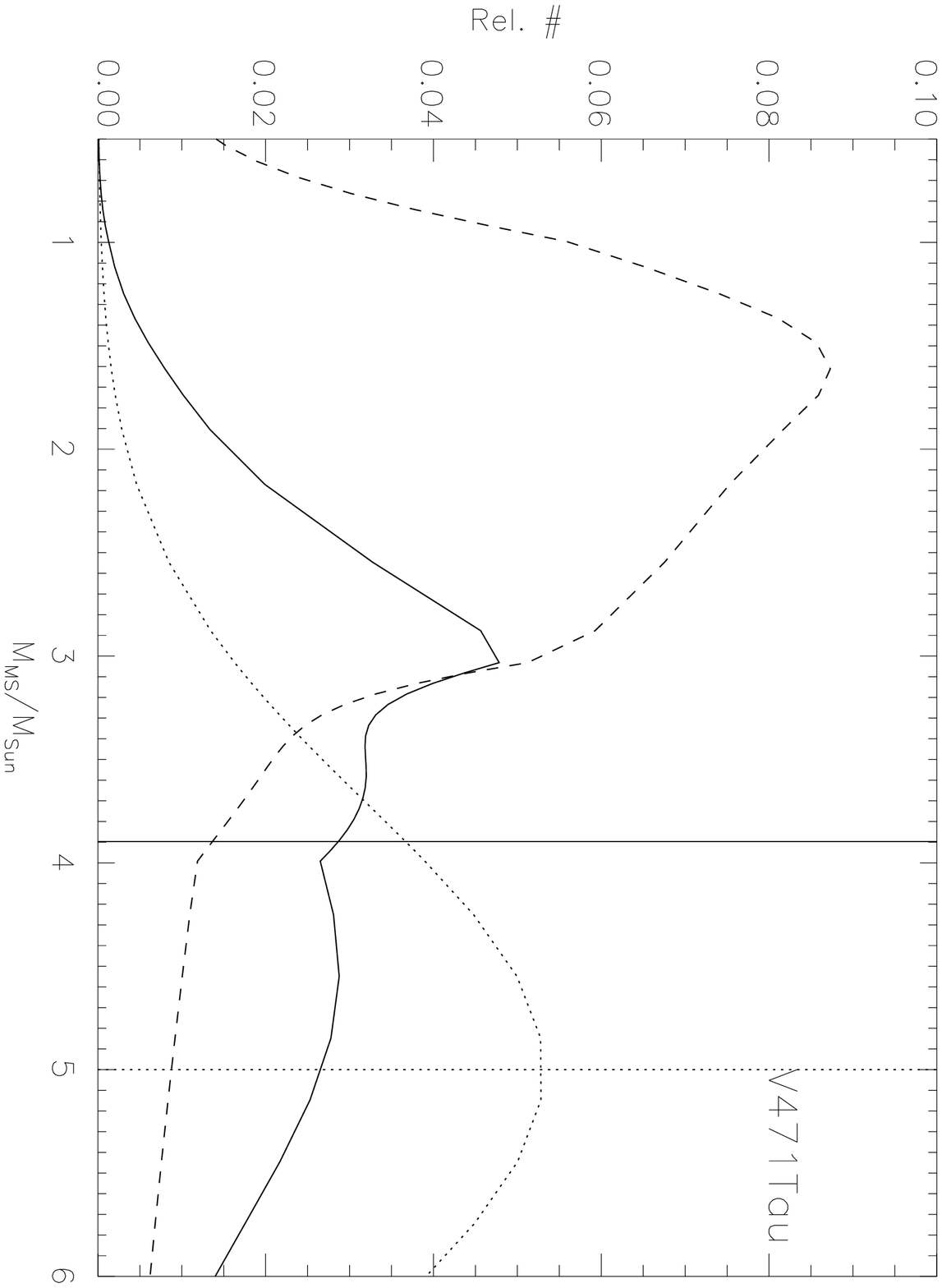}
 	\end{center}\caption{The main sequence mass (solid vertical lines) determined for two of our post-CE primaries: A~63, a central star of PN (left) and V471~Tau, a post-AGB star (right). These masses are the means derived from statistical distributions (solid curves), that are the result of a convolution of the Gaussian mass obtained from the IFMR (dotted curve with the mean plotted as dotted vertical lines) and the Baysian prior distributions (dashed lines). These latter distributions are the mass distributions of the progenitors of the two types of objects, the central stars of PN (left) and the post-AGB stars with no PN (right).}
	\label{fig:prior}
\end{figure*}

The last step in the main sequence mass determination is carried out by using conditional probability considerations. For each of the three populations of post-CE binaries (the central stars of PN, the post-AGB stars with no PN and the post-RGB stars) we know the main sequence mass distribution of the progenitors from population considerations and observations (see below).  To exploit this extra knowledge, we convert the progenitor main sequence mass of each system's primary, determined above from the WD mass alone (dotted vertical lines in Fig.~\ref{fig:prior}), into a Gaussian probability distribution function (PDF), by using the error estimates (dotted curves in Fig.~\ref{fig:prior}). We then multiply each PDF with the main sequence mass distribution for the parent population (also called the prior in Bayesian statistics; dashed curves in Fig.~\ref{fig:prior}). This procedure results in a PDF for the ``conditional" values of the main sequence masses of each of our post-CE primaries (solid curves in Fig.~\ref{fig:prior}), from which we can determine a mean value (solid vertical lines in Fig.~\ref{fig:prior}) and a new error estimate. These are the values we adopt for the primary progenitors' main sequence masses. We note that two identical primary masses ($M_c$) that have different error estimates will result in two different ``conditional" main sequence progenitor masses, because of the Bayesian statistical treatment (compare HW~Vir, HS0705+6700 and MS~Peg in Table~\ref{tab:alphainputs}). In Fig.~\ref{fig:prior} we show two specific cases, one for a central stars of PN, and one for a post-AGB system with no PN.

In Fig.~\ref{fig:prior} (dashed lines) the prior distributions are the main sequence mass distribution of the parent populations. For the central stars of PN,  the progenitor main sequence mass distribution is known from population synthesis as well as observations \citep[][dashed line in the left panel of Fig.~\ref{fig:prior}]{Moe2006}. For the post-AGB stars with no PN, the main sequence mass distribution of the progenitors can be determined using the WD mass distribution \citep{Kepler2007} and using the IFMR to translate the WD mass distribution into a main sequence mass distribution. The WD mass distribution of \citet{Kepler2007} is fitted by three Gaussian curves with different means and standard deviations. The Gaussian curve with the lowest mean mass corresponds to the helium WDs and we ignore it. To represent the WD mass distribution corresponding to the post-AGB stars (with no PN), we only use the remaining two Gaussian curves combined (with mean masses of 0.578 and 0.678~\msun,  respectively, standard deviations 0.047 and 0.148~\msun, respectively and relative strength 2.9:1)\footnote{We note that this prior is not entirely correct, because CE interactions on the AGB happen slightly more frequently to relatively more massive stars. This is because lower mass stars grow to relatively larger radii while on the RGB, increasing the chance that an interaction will happen on the RGB rather than the AGB. As a result the priors for the post-AGB, post-CE binary populations should be slightly weighted towards more massive stars. To calculate the correct priors would entail a complete population study, which we have not carried out here.}. 
%We used these main sequence mass distributions as the ``prior" probability distribution functions for the progenitors of the central stars and post-AGB groups of primary stars in our post-CE binaries. 

For the post-RGB group we cannot use the IFMR, because the primaries have not been through the AGB evolution. For all post-RGB primaries, we therefore use the initial mass function ($\propto M^{-2.35}$; \citealt{Kroupa2001}), truncated between 0.78 and 2.3~\msun\ \citep{Nelemans2000}. This distribution results in a mean value accompanied by an error (1.19$\pm$0.40~\msun).  

Once we have calculated the primaries' progenitors main sequence masses, we can derive the mass of the primary at the time of the CE interaction ($M_1$), by determining how much mass the star lost between the main sequence and the time of the CE interaction on the RGB or AGB. The stellar evolutionary calculations of \citet{Bertelli2008} for a range of masses and metallicities are used once again to average the RGB and AGB stellar masses using uniform weighting with respect to radius (for the AGB, we used only the values for which the radius was larger than the maximum radius attained on the RGB). The values thus obtained are $M_1/M_{\rm MS}$ = 0.98$\pm$0.02 for post-RGB stars and 0.87$\pm$0.07 for post-AGB stars, where the errors are derived from the scatter on the fit. We then considered that the presence of a companion would stimulate mass-loss prior to the CE interaction \citep[e.g.,][]{Bear2010}, so we lowered these numbers slightly and raised the uncertainty: $M_1/M_{\rm MS}$ = 0.90$\pm$0.10 for RGB systems and $0.75 \pm 0.15$ for AGB ones. Once again, the dependence on the mass and metallicity is not what dominates the error, but rather the lack of knowledge of what stage during the RGB or AGB evolution the systems enter the CE phase.

\subsubsection{The primary radius at the time of the CE interaction}
 
\begin{figure*}
\vspace{6cm}
	\begin{center}
		\includegraphics{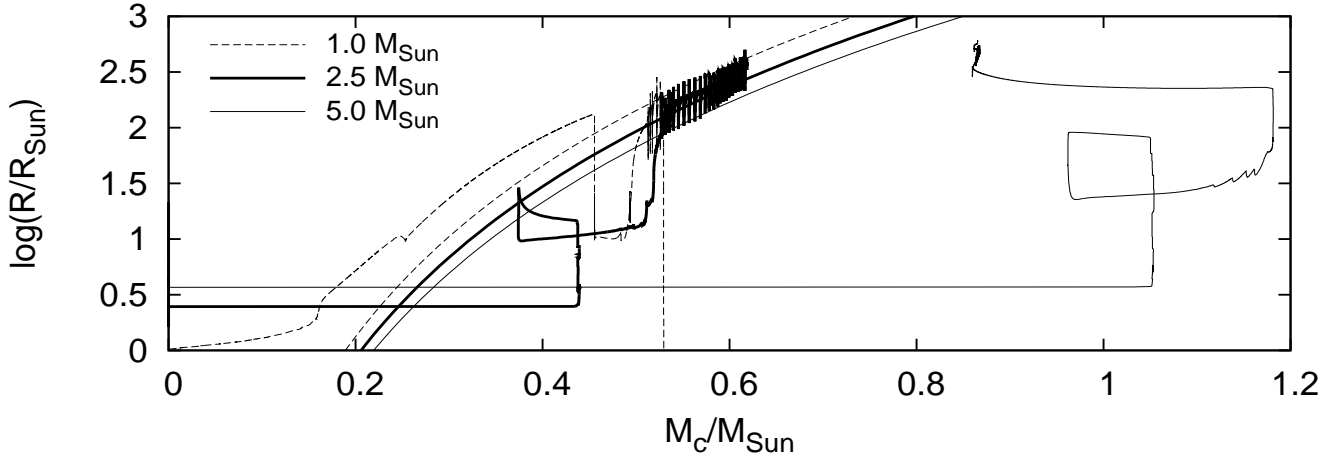}
	\end{center}\caption{Stellar evolutionary tracks on the core mass vs. stellar radius plane for 1~\msun, 2.5~\msun\ and 5~\msun (thick solid, dashed and thin solid lines, respectively) calculated with the MESA code, accompanied by fits to the \citet{Bertelli2008} tracks (smooth curves, Eq.~\ref{eq:radiusvsMms}) for the same three masses.}
	\label{fig:tracks}
\end{figure*}

We next determined the radii of the giants at the time of the CE interaction. From the MESA calculations we know the radius evolution as a function of the mass of the hydrogen-exhausted core (Fig.~\ref{fig:tracks}). We first selected the track corresponding to the main sequence progenitor's mass of each of our post-CE primary stars (\S~\ref{sssec:mass}); we then read from the plot the radius corresponding to today's core mass. This is the radius of the giant at the time of the CE interaction. The error on this radius has to be determined taking into account the errors for both the measured core mass {\it and} the derived main sequence mass. Considering each error bar in turn results in new radius values, from which we can estimate the final error on the radius. There is, however, one additional complication. The radius evolution of each model is complex and non-monotonic. This means that we need to use logical arguments to consider each radius value obtained in this way: a star can only be caught in a CE interaction while its radius is growing {\it and} if it is larger than at any time in the past. Once all these conditions are accounted for, it is found that the range of possible radii corresponding to each of our systems is quite large and that it is no less accurate to use fits of the core mass vs. radius values from the \citet{Bertelli2008} calculations, for the usual ranges of masses and metallicities: 

\begin{equation}
R = 440 R_\odot  (M_{\rm MS}/M_\odot)^{-0.47} (M_c/(0.6 M_\odot))^{5.1} (Z/Z_\odot)^{0.15},
\label{eq:radiusvsMms}
\end{equation}

\noindent where the {\it rms} error on the fit is $\pm 50$~\rsun. Or, evaluating $R$ in terms of $M_1$ instead of $M_{\rm MS}$:

\begin{equation}
R = 440 R_\odot  (M_1/M_\odot)^{-0.54}  (M_c/(0.6 M_\odot))^{5.0} (Z/Z_\odot)^{0.14} ,
\label{eq:radius}
\end{equation}

\noindent with the same fit scatter. We note that this relation applies equally to the post-RGB and post-AGB stars, although the fit was technically carried out for the AGB phase only. These fits are over-plotted on the detailed MESA calculation in Fig.~\ref{fig:tracks}. As can be seen from Fig.~\ref{fig:tracks}, the radius values implied by our approximate fits and the range of radii one would derive using the detailed MESA tracks become quite different for larger main sequence masses. The most massive of our systems, V~471~Tau, with a core mass of 0.84~\msun, has a conditional main sequence mass of 3.9~\msun. Using the MESA tracks one derives a radius at the time of the CE interaction of 435$\pm$100~\rsun. The value derived instead from the fits is 570$\pm$62 (see Table~\ref{tab:alphainputs}). These two values are quite different, although they are consistent within the uncertainties. Pre-emptying our results from \S~\ref{ssec:results}, the MESA radius value results in \ace=0.28$\pm$0.16, while using the radius fit from Eq.~\ref{eq:radius} \ace=0.21$\pm$0.15. We therefore see that even for the worse case scenario of the system with the largest mass, the advantage of using the detailed MESA code for this purpose is limited.

Finally, as we did for the case of the main sequence mass, we combine these radius values and their formal errors with the prior knowledge that for our primaries' progenitor populations, the radii of the RGB primaries should be in the range 10--300~\rsun, while for the AGB primaries they should be in the range 50--650~\rsun \citep{Bertelli2008}, in the usual mass and metallicity ranges. The maximum RGB radius of 300~\rsun, and maximum AGB radius of 650~\rsun, are the maximum radii achieved by any model in the set considered. The minimum radius value adopted for the AGB (50~\rsun) was selected because it is the smallest of all maximum RGB radii, achieved by any of the considered models. The minimum RGB radius (10~\rsun) is 
the approximate radius that separates Hertzsprung gap from RGB models for the \citet{Bertelli2008} tracks.  Convolution with the prior results in a new radius value and a new error, according to Bayesian statistics. 

\subsection{Observed systems used in the \\ determination of \ace}%\\ Obtaining $M_1$ and $R_i$ from $M_c$}
\label{ssec:obsandalpha}

In Table~\ref{tab:alphainputs} we list the values of $M_c$, $M_2$, $P$ and $A_f$ determined from observations as well as the values of $M_1$ and $R$ deduced from stellar evolution in \S~\ref{ssec:reconstruction}. References for the observed values can be obtained  from \citet{DeMarco2009b}, for central stars of PN. Of all the $\sim$40 post-CE binary central stars listed by \citet{DeMarco2009b} only 6 central stars with non-degenerate secondaries have sufficient information to be useful in the present study. For the observed post-CE binary in the centre of PN NGC~6337 we list two sets of parameters in Tables~\ref{tab:alphainputs} and \ref{tab:alphaoutputs}. When we fit the value of \ace\  (\S~\ref{sssec:anti-correlation}), we use only one set (that with $M_2=0.35$~\msun). The fit with the alternative set, results in very similar fit parameters. Finally, for the binary system in the PN HFG~1 there also are two sets of parameters \citep{DeMarco2009b}. We use the set with the most massive companion, $M_2=1.09$~\msun, because a recent analysis of X-ray radiation \citep{Montez2010} finds this system's companion to be coronally active and the only way to reproduce the X-ray luminosity is to have a companion at least as massive as a $\sim 1$~\msun\ main sequence star.

Other post-CE binaries are primarily from the list of  \citet[][where we selected only those systems that have not suffered substantial reduction of the orbital period after emerging from the CE]{Schreiber2003}. Wherever parameters have been updated by new measurements of \ZT, we have adopted the revisions. From the extensive list of {\it Sloan Digital Sky Survey} post-CE binaries in \ZT, we only use the systems SDSS~J1548+4057, SDSS~J0110+1326 and SDSS~J1435+3733 (for more discussion, see \S~\ref{sec:summary}). Finally, we have also considered AA~Dor \citep{Mueller2010} and HD~149382 \citep{Geier2009}, because of their low mass secondaries. Whenever errors were not given in the cited references, we have assumed an error of 20\% on both primary and secondary masses. In Table~\ref{tab:alphaoutputs} we list the calculated values of $q$ (=$M_2/M_1$) and  \ace, determined using Eq.~\ref{eq:alphaformalism}, along with their uncertainties, both in linear and logarithmic form. We also list values of \ace\ determined using the \citet{Webbink2008} relation.

\begin{table*}
\begin{tabular}{llccccccc} \hline
Name & Type$^1$ & $M_c$ & $M_2$ & P & $A_f$ & $M_{MS}$ & $M_1$ & $R$\\
  &   &  (\msun) & (\msun) & (days) & (\rsun) & (\msun) & (\msun) & (\rsun)\\ \hline
          BE~UMa      & CSPN &   0.70$\pm$  0.07  &   0.36$\pm$  0.07  & 2.29 &    7.5$\pm$     0.7  &    3.4$\pm$     0.4  &    2.5$\pm$     0.6  &      440$\pm$     131 \\
            A~46      & CSPN &   0.51$\pm$  0.07  &   0.15$\pm$  0.02  & 0.47 &    2.2$\pm$     0.2  &    1.2$\pm$     0.3  &    0.90$\pm$     0.29  &      239$\pm$     140 \\
            A~63      & CSPN &   0.63$\pm$  0.06  &   0.29$\pm$  0.04  & 0.46 &    2.4$\pm$     0.2  &    2.4$\pm$     0.9  &    1.8$\pm$     0.8  &      376$\pm$     140 \\
           HFG~1      & CSPN &   0.57$\pm$  0.10  &   1.1$\pm$  0.1  & 0.58 &    3.5$\pm$     0.4  &    1.4$\pm$     0.5  &    1.1$\pm$     0.4  &      316$\pm$     172 \\
            DS~1      & CSPN &   0.63$\pm$  0.03  &   0.23$\pm$  0.01  & 0.36 &    2.0$\pm$     0.1  &    3.0$\pm$     0.4  &    2.2$\pm$     0.5  &      348$\pm$      93 \\
        NGC~6337      & CSPN &   0.60$\pm$  0.07  &   0.35$\pm$  0.04  & 0.17 &    1.3$\pm$     0.1  &    1.8$\pm$     0.8  &    1.4$\pm$     0.7  &      346$\pm$     152 \\
&&&(0.20$\pm$0.03)&&(1.2$\pm$0.1)&&&\\
        V471~Tau      & pAGB &   0.84$\pm$  0.05  &   0.93$\pm$  0.10  & 0.52 &    3.3$\pm$     0.2  &    3.9$\pm$     1.3  &    2.9$\pm$     1.1  &      570$\pm$      62 \\
          UZ~Sex      & pAGB &   0.65$\pm$  0.23  &   0.22$\pm$  0.05  & 0.60 &    2.8$\pm$     0.8  &    2.4$\pm$     1.1  &    1.8$\pm$     0.9  &      322$\pm$     194 \\
  SDSS~J1548+4057      & pAGB &   0.65$\pm$  0.03  &   0.17$\pm$  0.03  & 0.19 &    1.3$\pm$     0.1  &    3.1$\pm$     0.2  &    2.3$\pm$     0.5  &      383$\pm$      99 \\
    RE~J1016-053      & pAGB &   0.60$\pm$  0.02  &   0.15$\pm$  0.02  & 0.79 &    3.3$\pm$     0.1  &    2.6$\pm$     0.5  &    1.9$\pm$     0.6  &      288$\pm$      66 \\
        Feige~24      & pAGB &   0.57$\pm$  0.03  &   0.39$\pm$  0.02  & 4.23 &   10.9$\pm$     0.4  &    1.9$\pm$     0.6  &    1.4$\pm$     0.5  &      266$\pm$      85 \\
          IN~CMa      & pAGB &   0.57$\pm$  0.03  &   0.43$\pm$  0.03  & 1.26 &    4.9$\pm$     0.2  &    1.9$\pm$     0.6  &    1.4$\pm$     0.5  &      266$\pm$      85 \\
    RE~J2013+400      & pAGB &   0.56$\pm$  0.03  &   0.18$\pm$  0.04  & 0.71 &    3.0$\pm$     0.2  &    1.8$\pm$     0.6  &    1.4$\pm$     0.5  &      250$\pm$      81 \\
          NN~Ser      & pAGB &   0.53$\pm$  0.01  &   0.12$\pm$  0.03  & 0.13 &    0.94$\pm$     0.05  &    1.4$\pm$     0.3  &    1.1$\pm$     0.3  &      199$\pm$      38 \\
          GK~Vir      & pAGB &   0.51$\pm$  0.04  &   0.10$\pm$  0.01  & 0.34 &    1.7$\pm$     0.1  &    1.3$\pm$     0.4  &    1.0$\pm$     0.3  &      197$\pm$      84 \\
            Hz~9      & pAGB &   0.51$\pm$  0.10  &   0.28$\pm$  0.04  & 0.56 &    2.6$\pm$     0.4  &    1.7$\pm$     0.7  &    1.2$\pm$     0.5  &      245$\pm$     161 \\
          AA~Dor      & pAGB &   0.51$\pm$  0.12  &   0.085$\pm$  0.027  & 0.26 &    1.4$\pm$     0.3  &    1.7$\pm$     0.7  &    1.3$\pm$     0.6  &      256$\pm$     172 \\
        BPM~6502      & pAGB &   0.50$\pm$  0.05  &   0.17$\pm$  0.01  & 0.34 &    1.8$\pm$     0.1  &    1.2$\pm$     0.4  &    0.90$\pm$     0.32  &      198$\pm$     101 \\
     PG~1017-086      & pAGB &   0.50$\pm$  0.05  &   0.078$\pm$  0.006  & 0.07 &    0.64$\pm$     0.05  &    1.2$\pm$     0.4  &    0.90$\pm$     0.32  &      198$\pm$     101 \\
          NY~Vir      & pAGB &   0.50$\pm$  0.05  &   0.15$\pm$  0.02  & 0.10 &    0.79$\pm$     0.07  &    1.2$\pm$     0.4  &    0.90$\pm$     0.32  &      198$\pm$     101 \\
  SDSS~J1435+3733      & pAGB &   0.50$\pm$  0.03  &   0.22$\pm$  0.03  & 0.13 &    0.95$\pm$     0.05  &    1.1$\pm$     0.3  &    0.85$\pm$     0.27  &      181$\pm$      54 \\
          HW~Vir      & pAGB &   0.48$\pm$  0.09  &   0.14$\pm$  0.02  & 0.12 &    0.86$\pm$     0.13  &    1.2$\pm$     0.4  &    0.90$\pm$     0.36  &      224$\pm$     150 \\
    HS~0705+6700      & pAGB &   0.48$\pm$  0.05  &   0.13$\pm$  0.03  & 0.10 &    0.75$\pm$     0.07  &    1.0$\pm$     0.3  &    0.74$\pm$     0.25  &      181$\pm$      95 \\
          MS~Peg      & pAGB &   0.48$\pm$  0.02  &   0.22$\pm$  0.02  & 0.17 &    1.1$\pm$     0.0  &    0.88$\pm$     0.14  &    0.66$\pm$     0.17  &      154$\pm$      39 \\
  SDSS~J0110+1326      & pAGB &   0.47$\pm$  0.02  &   0.31$\pm$  0.05  & 0.33 &    1.9$\pm$     0.1  &    0.85$\pm$     0.13  &    0.64$\pm$     0.16  &      141$\pm$      36 \\
          LM~Com      & pRGB &   0.45$\pm$  0.05  &   0.28$\pm$  0.05  & 0.26 &    1.5$\pm$     0.1  &    1.2$\pm$     0.4  &    1.1$\pm$     0.4  &      116$\pm$      61 \\
          RR~Cae      & pRGB &   0.44$\pm$  0.02  &   0.18$\pm$  0.01  & 0.30 &    1.6$\pm$     0.1  &    1.2$\pm$     0.4  &    1.1$\pm$     0.4  &       89$\pm$      29 \\
       HD~149382      & pRGB &   0.47$\pm$  0.12  &   0.015$\pm$  0.007  & 2.39 &    5.9$\pm$     1.5  &    1.2$\pm$     0.4  &    1.1$\pm$     0.4  &      136$\pm$      91 \\
          HR~Cam      & pRGB &   0.41$\pm$  0.01  &   0.10$\pm$  0.01  & 0.10 &    0.72$\pm$     0.02  &    1.2$\pm$     0.4  &    1.1$\pm$     0.4  &       59$\pm$      13 \\
          CC~Cet      & pRGB &   0.39$\pm$  0.10  &   0.18$\pm$  0.05  & 0.28 &    1.5$\pm$     0.3  &    1.2$\pm$     0.4  &    1.1$\pm$     0.4  &       96$\pm$      82 \\
     WD~0137-349      & pRGB &   0.39$\pm$  0.04  &   0.052$\pm$  0.005  & 0.08 &    0.60$\pm$     0.05  &    1.2$\pm$     0.4  &    1.1$\pm$     0.4  &       54$\pm$      26 \\
           DeMa $^2$ &  Sim &   0.60    &  0.10  &  30    & 36 &  1.20 &    1.0    &   645 \\
        Sand1-2 $^2$ &  Sim &   0.70    &  0.40  &   1.08    &  4.56 &  3.18 &    2.8    &   190 \\
          Sand3 $^2$ &  Sim &   1.00    &  0.40  &   0.90    &  4.39 &  5.29 &    4.6    &   190 \\
          Sand4 $^2$ &  Sim &   1.00    &  0.60  &   0.96    &  4.79 &  5.29 &    4.6    &   190 \\
          Sand5 $^2$ &  Sim &   0.94    &  0.60  &   2.48    &  8.90 &  5.31 &    4.6    &   353 \\
\hline \multicolumn{9}{l}{$^1$CSPN: the primary is a central star of PN and went through the AGB evolution; pAGB: the primary}\\ \multicolumn{9}{l}{does not have a PN, but we consider it as a pot-AGB star; pRGB: the primary is considered as having suffered a}\\ \multicolumn{9}{l}{CE interaction on the RGB and never having ascended the AGB; Sim: simulations.}\\ \multicolumn{9}{l}{$^2$DeMa: \citet{DeMarco2003}. Sand: \citet{Sandquist1998}, simulations. The number indicates the particular simulation,}\\ \multicolumn{9}{l}{and follows the scheme of their Table 1. See also text in \S~\ref{ssec:simsandalpha}.} \end{tabular}

\caption{The observationally-derived quantities for our post-CE systems ($M_c$, $M_2$ and $P$) listed alongside the quantities derived from orbital, stellar evolution and population considerations ($A_f$, $M_1$ and $R$). Simulation inputs ($M_c$, $M_2$, $M_1$ and $R$) and outputs ($P$ and $A_f$) are also listed. All these quantities are inputs to the \ace\ equation (Eq.~\ref{eq:alphaformalism}). Parameters in brackets are an alternative set not used in the fits (see \S~\ref{ssec:obsandalpha}) \label{tab:alphainputs}}
\end{table*}

\begin{table*}
\begin{tabular}{llccccc} \hline
Name & Type$^1$ & $\log q$ & $\log \alpha$ & $q$ & $\alpha$ & $\alpha$(Web)  \\ \hline
        BE~UMa      & CSPN &  -0.85$\pm$  0.12 &   0.08$\pm$  0.25 &     0.14$\pm$    0.04  &     1.2$\pm$    0.7     &      1.1   \\
          A~46      & CSPN &  -0.78$\pm$  0.13 &  -0.45$\pm$  0.34 &     0.17$\pm$    0.05  &     0.35$\pm$    0.31     &      0.24   \\
          A~63      & CSPN &  -0.78$\pm$  0.17 &  -0.54$\pm$  0.34 &     0.16$\pm$    0.06  &     0.29$\pm$    0.25     &      0.23   \\
         HFG~1      & CSPN &   0.01$\pm$  0.16 &  -1.06$\pm$  0.35 &     1.0$\pm$    0.4  &     0.087$\pm$    0.080     &      0.043   \\
          DS~1      & CSPN &  -0.99$\pm$  0.10 &  -0.29$\pm$  0.23 &     0.10$\pm$    0.02  &     0.51$\pm$    0.28     &      0.46   \\
      NGC~6337      & CSPN &  -0.59$\pm$  0.18 &  -1.01$\pm$  0.36 &     0.26$\pm$    0.11  &     0.10$\pm$    0.09     &      0.068   \\
                             &          &(-0.83$\pm$ 0.19) & (-0.83$\pm$0.34) &       (0.15$\pm$0.06)      &     (0.15$\pm$0.14)        &      (0.11)    \\
      V471~Tau      & pAGB &  -0.50$\pm$  0.15 &  -0.68$\pm$  0.30 &     0.32$\pm$    0.11  &     0.21$\pm$    0.15     &      0.17   \\
        UZ~Sex      & pAGB &  -0.91$\pm$  0.20 &  -0.24$\pm$  0.45 &     0.12$\pm$    0.06  &     0.57$\pm$    0.70     &      0.37   \\
SDSS~J1548+4057      & pAGB &  -1.13$\pm$  0.10 &  -0.32$\pm$  0.22 &     0.074$\pm$    0.018  &     0.48$\pm$    0.26     &      0.31   \\
  RE~J1016-053      & pAGB &  -1.11$\pm$  0.12 &   0.13$\pm$  0.25 &     0.078$\pm$    0.023  &     1.3$\pm$    0.8     &      0.85   \\
      Feige~24      & pAGB &  -0.56$\pm$  0.14 &   0.08$\pm$  0.29 &     0.27$\pm$    0.09  &     1.2$\pm$    0.9     &      0.86   \\
        IN~CMa      & pAGB &  -0.52$\pm$  0.14 &  -0.32$\pm$  0.30 &     0.30$\pm$    0.10  &     0.47$\pm$    0.35     &      0.35   \\
  RE~J2013+400      & pAGB &  -0.88$\pm$  0.16 &  -0.16$\pm$  0.31 &     0.13$\pm$    0.05  &     0.69$\pm$    0.53     &      0.44   \\
        NN~Ser      & pAGB &  -0.94$\pm$  0.15 &  -0.54$\pm$  0.25 &     0.11$\pm$    0.04  &     0.29$\pm$    0.18     &      0.17   \\
        GK~Vir      & pAGB &  -0.98$\pm$  0.13 &  -0.22$\pm$  0.30 &     0.10$\pm$    0.03  &     0.60$\pm$    0.45     &      0.34   \\
          Hz~9      & pAGB &  -0.65$\pm$  0.17 &  -0.40$\pm$  0.41 &     0.22$\pm$    0.09  &     0.40$\pm$    0.43     &      0.28   \\
        AA~Dor      & pAGB &  -1.17$\pm$  0.20 &  -0.16$\pm$  0.44 &     0.07$\pm$    0.03  &     0.69$\pm$    0.81     &      0.40   \\
      BPM~6502      & pAGB &  -0.72$\pm$  0.13 &  -0.47$\pm$  0.33 &     0.19$\pm$    0.06  &     0.34$\pm$    0.28     &      0.21   \\
   PG~1017-086      & pAGB &  -1.06$\pm$  0.14 &  -0.60$\pm$  0.33 &     0.087$\pm$    0.027  &     0.25$\pm$    0.21     &      0.14   \\
        NY~Vir      & pAGB &  -0.78$\pm$  0.14 &  -0.78$\pm$  0.33 &     0.17$\pm$    0.06  &     0.17$\pm$    0.14     &      0.10   \\
SDSS~J1435+3733      & pAGB &  -0.59$\pm$  0.13 &  -0.87$\pm$  0.27 &     0.26$\pm$    0.08  &     0.13$\pm$    0.09     &      0.084   \\
        HW~Vir      & pAGB &  -0.81$\pm$  0.16 &  -0.73$\pm$  0.40 &     0.16$\pm$    0.06  &     0.19$\pm$    0.20     &      0.11   \\
  HS~0705+6700      & pAGB &  -0.76$\pm$  0.15 &  -0.82$\pm$  0.34 &     0.18$\pm$    0.06  &     0.15$\pm$    0.13     &      0.087   \\
        MS~Peg      & pAGB &  -0.48$\pm$  0.11 &  -0.91$\pm$  0.27 &     0.33$\pm$    0.08  &     0.12$\pm$    0.08     &      0.074   \\
SDSS~J0110+1326      & pAGB &  -0.31$\pm$  0.12 &  -0.82$\pm$  0.28 &     0.49$\pm$    0.13  &     0.15$\pm$    0.11     &      0.10   \\
        LM~Com      & pRGB &  -0.58$\pm$  0.15 &  -0.43$\pm$  0.34 &     0.26$\pm$    0.09  &     0.37$\pm$    0.32     &      0.26   \\
        RR~Cae      & pRGB &  -0.77$\pm$  0.14 &  -0.09$\pm$  0.29 &     0.17$\pm$    0.05  &     0.81$\pm$    0.57     &      0.54   \\
     HD~149382      & pRGB &  -1.85$\pm$  0.21 &   1.52$\pm$  0.47 &     0.014$\pm$    0.007  &    33$\pm$   43     &     16   \\
        HR~Cam      & pRGB &  -1.03$\pm$  0.14 &   0.04$\pm$  0.27 &     0.093$\pm$    0.030  &     1.1$\pm$    0.7     &      0.68   \\
        CC~Cet      & pRGB &  -0.77$\pm$  0.17 &  -0.09$\pm$  0.45 &     0.17$\pm$    0.07  &     0.82$\pm$    1.02     &      0.56   \\
   WD~0137-349      & pRGB &  -1.31$\pm$  0.14 &   0.30$\pm$  0.33 &     0.049$\pm$    0.016  &     2.0$\pm$    1.7     &      1.2   \\
         DeMa $^2$ &  Sim &  -1.02    &  0.53 &   0.096    &  3.4 &     1.8   \\
      Sand1-2 $^2$ &  Sim &  -0.84    &  0.34 &   0.14    &  2.2 &     1.6   \\
        Sand3 $^2$ &  Sim &  -1.06    &  0.96 &   0.087    &  9.0 &     6.3   \\
        Sand4 $^2$ &  Sim &  -0.88    &  0.82 &   0.13    &  6.6 &     4.8   \\
        Sand5 $^2$ &  Sim &  -0.89    &  0.86 &   0.13    &  7.2 &     5.4   \\
\hline \multicolumn{7}{l}{$^{1,2}$: see comments to Table~3} \end{tabular}

\caption{The determined values of \ace, using our equation as well as that of \citet{Webbink1984,Webbink2008}. Parameters in brackets are from an alternative sets of inputs (see Table~\ref{tab:alphainputs} and \S~\ref{ssec:obsandalpha}). \label{tab:alphaoutputs}}
\end{table*}
	
% =============================================================================
%							SIMS AND ALPHA
% =============================================================================

\subsection{The simulations used in the \\ determination of \ace}
\label{ssec:simsandalpha}

Hydrodynamic simulations, carried out both with smooth particle hydrodynamics \citep[e.g.][]{Terman1996} or grid methods \citep[e.g.,][]{Sandquist1998}, are a tool to determine the value of \ace\ {\it ab initio}. These simulations start by mapping a giant star into the simulation domain. The quantities describing the star (i.e., density, temperature, internal energy, etc.) are calculated using a one-dimensional stellar evolution model. The companion is represented by a point mass, as is the compact core of the giant. The physics included in the simulation is typically only the gravitational attraction of all the masses involved. The interaction timescale is short enough that radiation transport can be neglected (e.g., see discussion in Sandquist et al. 1998). The in-spiral of the companion is typically followed until the resolution of the grid becomes insufficient, which may or may not be at the termination of the CE interaction (i.e., when sufficient mass has become unbound from the system to dictate the collapse of the primary and its detachment from its Roche lobe). At the end of a simulation, \ace\ can be determined by using Eq.~\ref{eq:alphaformalism}. 

We determined \ace\ from 5 simulations, four from \citet{Sandquist1998} and one from \citet{DeMarco2003}. All simulations were carried out with the \citet{Burkert1993} code, as modified by \citet{Sandquist1998}. Relevant parameters are taken from the listed publications and are reproduced in Tables~\ref{tab:alphainputs} where, for the \citet{Sandquist1998} simulations, we use the same labels as in their Table~1. Their simulations 1 and 2 were identical, except for the giant envelope's rotation being either zero or synchronised with the companion's orbital motion. For these two simulations we take the average of their post-CE binary periods.  
 \begin{figure}
\vspace{16.5cm}
	\begin{center}
	\includegraphics{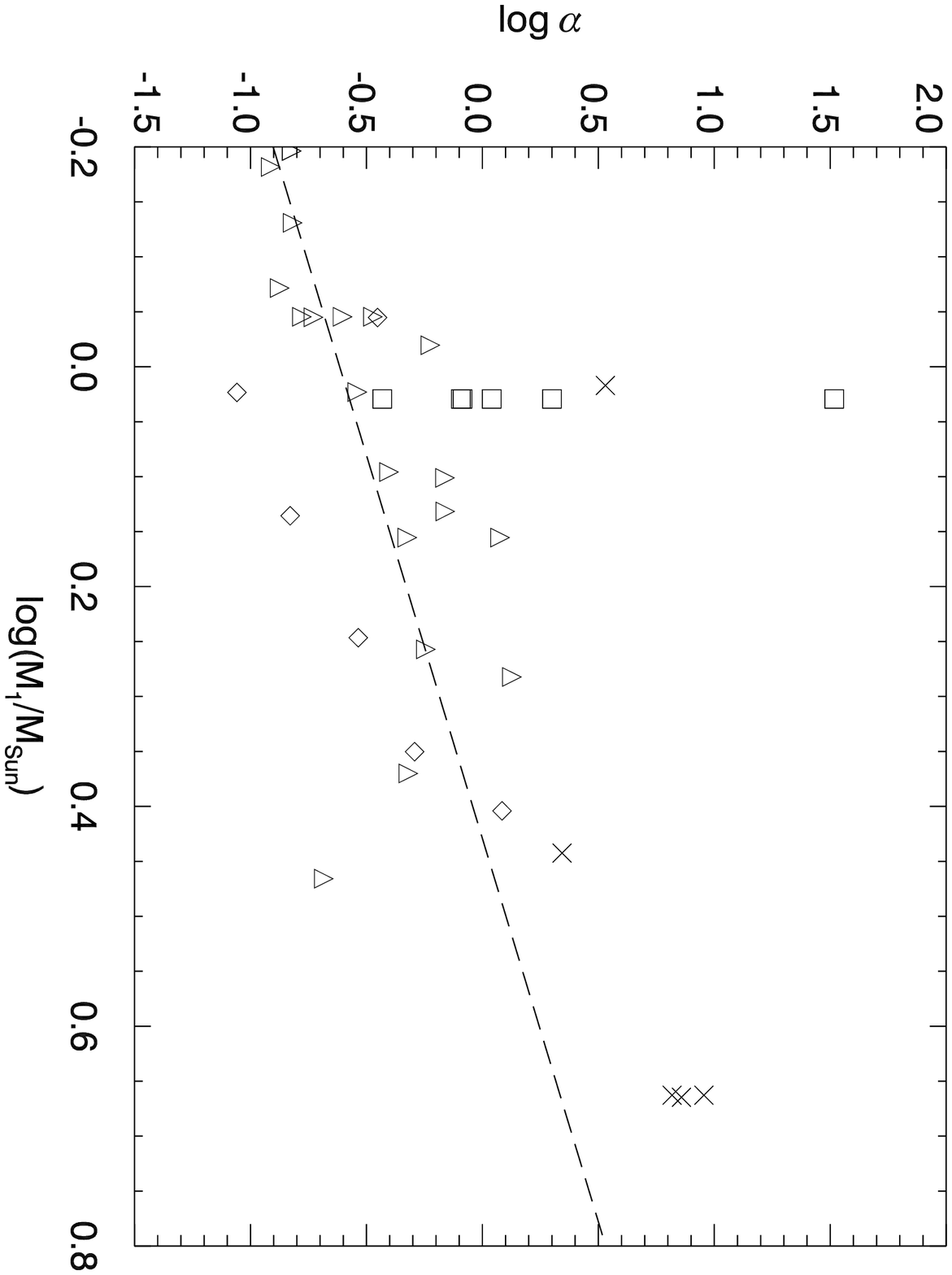}
	\includegraphics{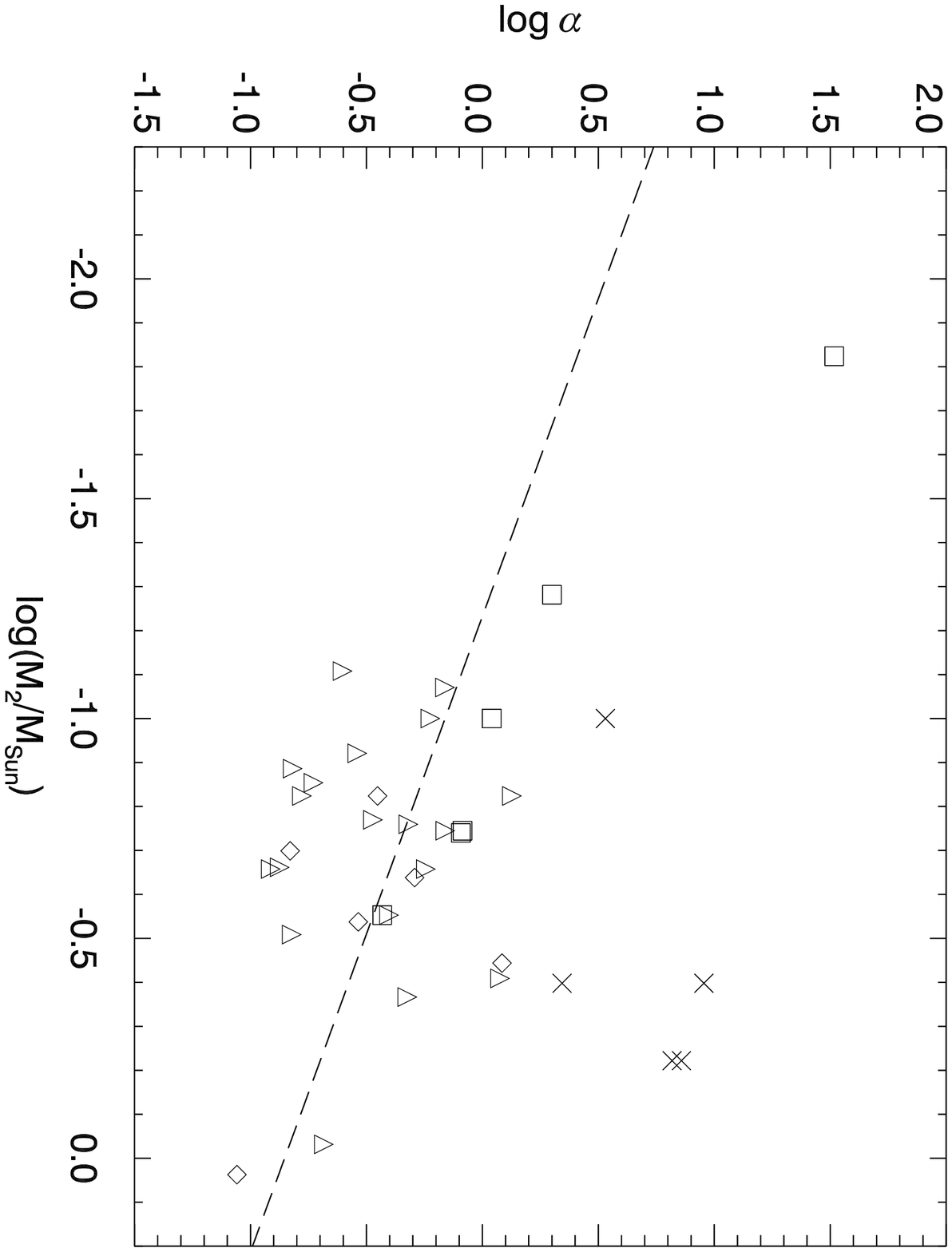}
	\includegraphics{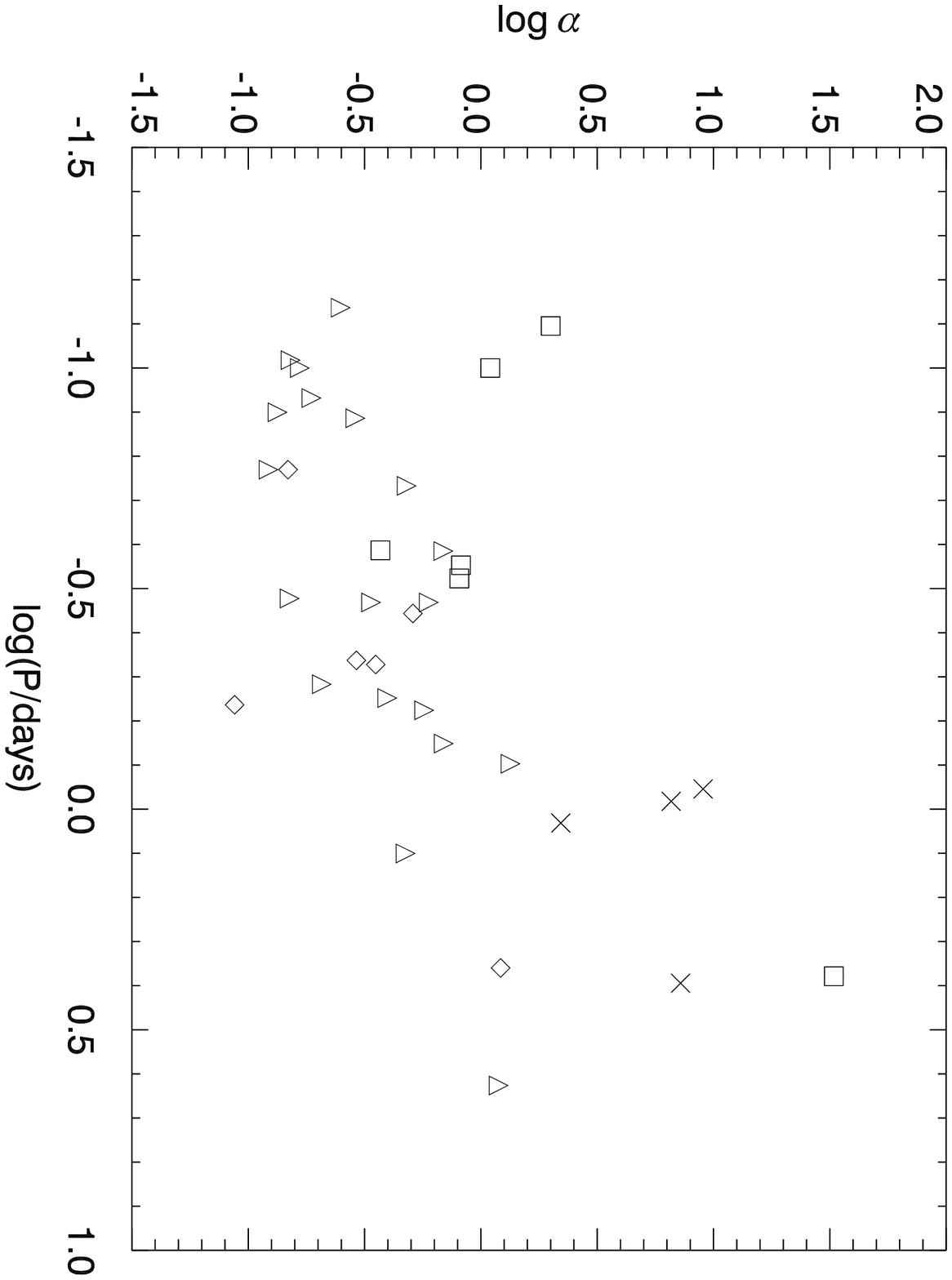}
 	\end{center}\caption{A plot of $\log$~\ace\ as a function of  $\log M_1$ (top),  $\log M_2$ (middle) and logarithm of the period (bottom). The symbols are the same as in Fig.~\ref{fig:alpha}. Error bars were omitted for clarity but were used in the linear fits (dashed lines). Post-RGB systems were not fitted in the upper panel. Simulations (crosses) were never fitted. }
	\label{fig:alpha2}
\end{figure}
All simulations considered were stopped after orbital decay time-scales lengthened considerably, when only a very small amount of envelope mass was left inside the orbit of the companion ($\sim 10^{-4}$ -- 10$^{-3}$~\msun). However, in all cases a considerable amount of envelope gas was still bound to the system albeit at some distance from the core (usually outside the original giant radius and with small binding energy). Because of this left over bound envelope, we may not assume that the separation of the binary at the end of the simulation is the actual post-CE binary separation. Any envelope still bound to the system at the end of the CE will fall back onto the system form a circumbinary disk that might have some dynamical effect on the binary period. To understand this eventuality, will require further numerical simulations. 

The simulations of \citet{Yorke1995} started the interaction with the companion well inside the AGB stellar envelope. This results in a negative value of \ace\ because $M_c / A_f < (M_c + M_e) / A_i$. Although they justified this initial setup by noting that spin-up only becomes important within their initial radius, when the timescale for orbital decay exceeds the orbital period, they also stated that a value for \ace\ cannot be trivially derived from their simulations, but that the values can be estimated between 0.3 and 0.6. Since it is not clear how these values were determined, we did not use their models.  

Finally, we did not use the simulations of \citet{Terman1996}. They considered a 5~\msun\ AGB giant in two different evolutionary stages.  The less evolved AGB star, still at the very base of the AGB, and still undergoing core helium burning, was used to simulate a CE interaction with a 0.5~\msun\ companion and presumably resulted in a merger. The more evolved of the two stars, was used to simulate CE interactions with 0.5 and 1.0~\msun\ main sequence companions. In neither case was the simulation followed till the end of the interaction, and a substantial amount of envelope remained bound to the system at the end of the simulation. However, the orbital decay time-scale had considerably slowed down and  envelope mass was still being unbound. The authors therefore concluded that the system would survive, albeit with a smaller orbital separation than that reached by the end of the simulation.

% =============================================================================
%							RESULTS
% =============================================================================

\subsection{Results}
\label{ssec:results}

Here, we finally calculate the values of \ace\ from observations and simulations (Table~\ref{tab:alphaoutputs}, where we report both linear and logarithmic values). The error bars on \ace\ are significant but realistic.

\subsubsection{\ace\ $>$ 1.} 
\label{sssec:alphagtunity}

For 6 observed systems and all the simulations \ace\ exceeds unity. The physical interpretation of this is that the specific CE interaction under consideration has benefited from more energy than the orbital energy supplied by the companion.
Several authors \citep[e.g.,][]{Han1995,Webbink2008} discuss thermal energy as well as the dissociation and ionisation energies as additional sources of energy in the \ace\ equation. In \S~\ref{ssec:virial} we discussed the thermal  energy and showed that we can use the virial theorem to gauge its value, which is exactly half of the binding energy. If we include the thermal energy in this way, all values of \ace\ will be below unity, except for the system with a substellar companion (HD~149382, discussed further in \S~\ref{sec:discussion}) and the simulations. 
%In \S~\ref{sec:discussion} we will speculate that if the the CE interaction takes place over longer than a stellar dynamical time, the thermal energy helps eject the envelope. 
 
\subsubsection{A possible $\log q$ vs. $\log$~\ace\ anti-correlation} 
\label{sssec:anti-correlation}

%NEW

In Fig.~\ref{fig:alpha2} we explore possible correlations between \ace\ and other relevant parameters. The simulations' \ace\ values are plotted (cross symbols)  but not fitted, because it is impossible to determine their error bars in a way that is consistent with the observations. We fit $\log M_1$ vs. $\log$\ace\ for post-AGB and CSPN systems (not for post-RGB primaries, because for those stars we had to adopt a single mass value). We find a very marginal correlation (correlation coefficient, $r$=0.56). On the $\log M_2$ vs. $\log$\ace\  plane we see a slightly better anti-correlation with $r = -0.79$ (post-RGB systems are fitted here). Fits to the data and their errors result in reduced $\chi^2$=0.77 and 1.79 for the $M_1$ and $M_2$ cases, respectively. The correlation between $\log$\ace\ and the period is extremely weak ($r = 0.33$).

%END NEW

\begin{table*}
\begin{tabular}{lcccccccc}
\hline
&&&\multicolumn{3}{c}{Linear fit}&\multicolumn{3}{c}{Constant fit}\\
        &     $n^a$ &  $r^a$  & $\log$ \ace = &    $\chi^2/\nu^b$ & pte$^a$ & \ace\ =& $\chi^2/\nu^c$ & pte$^a$ \\ \hline
All data                      & 31   & --0.78 & ($-1.4 \pm 0.2) - (1.2 \pm 0.2)\ \log q$  & 0.78 & 0.79  & 0.43$\pm$0.08 & 1.9 & 0\\
As above, no  HD~149382 & 30& --0.64& $(-1.2 \pm 0.2) - (1.0 \pm 0.3)\ \log q$  & 0.75 & 0.83  & 0.41$\pm$0.06 & 1.4 & 0.06\\
Only CSPN and pRGB   & 12& --0.92 & $(-1.4 \pm 0.3) - (1.4 \pm 0.3)\ \log q$  & 0.54 & 0.86  & 0.63$\pm$0.12 & 2.8 & 0\\
As above, no  HD~149382& 11& --0.83  & $(-1.2 \pm 0.3) - (1.2 \pm 0.4)\ \log q$  & 0.53 & 0.85 & 0.54$\pm$0.11 & 1.7 & 0\\

\hline
\multicolumn{9}{l}{$^a$$n$: size of the dataset. $r$: correlation coefficient. pte: probability to exceed.}\\
\multicolumn{9}{l}{$^b$Degrees of freedom, $\nu = n-2$.}\\
\multicolumn{9}{l}{$^c$Degrees of freedom, $\nu = n-1$.}\\
\end{tabular}
\caption{Statistical properties of the fits to $\log q$ vs. $\log$~\ace. \label{tab:statistics}}
\end{table*}

\begin{figure*}
\vspace{6.5cm}
	\begin{center}
	\includegraphics{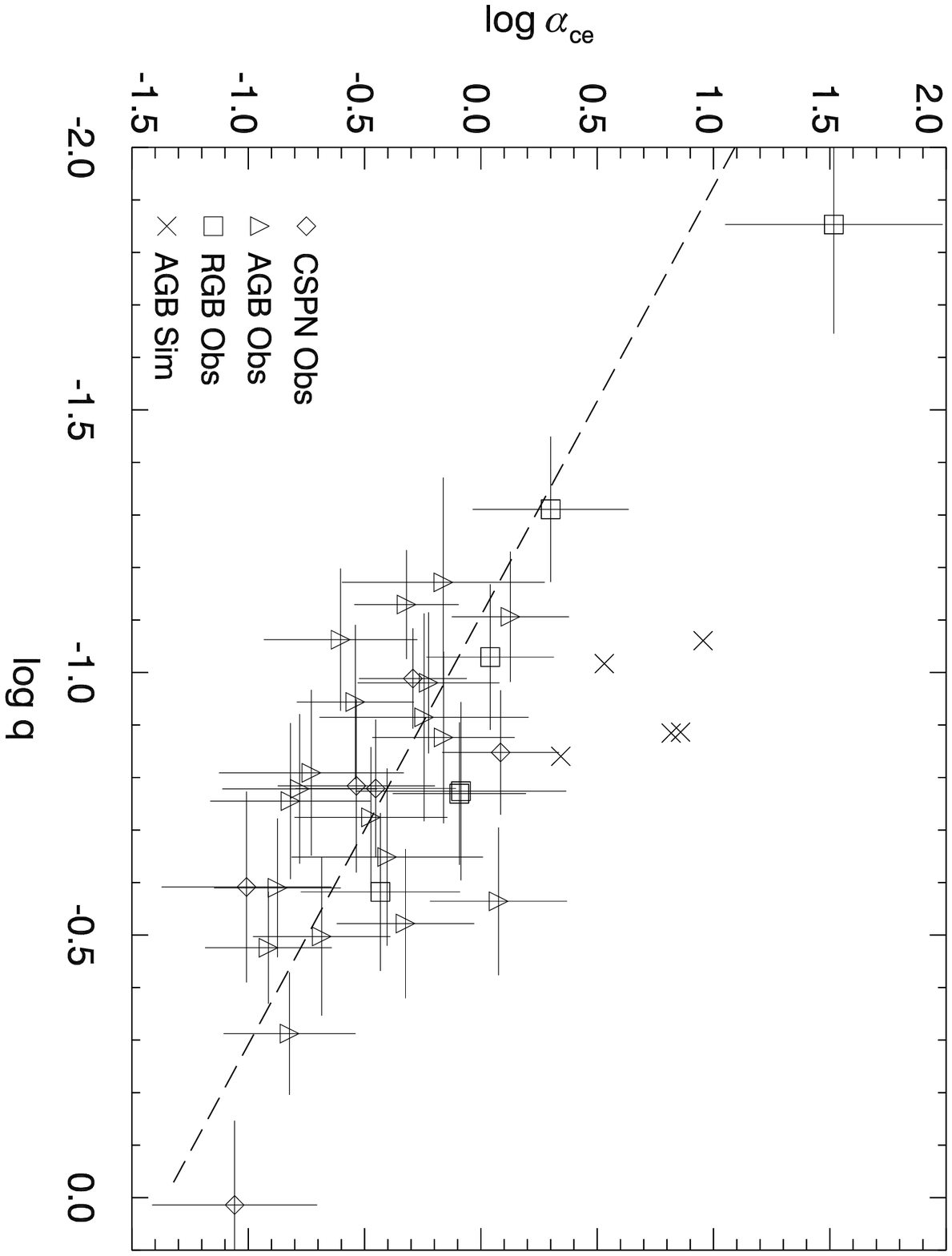}
	\includegraphics{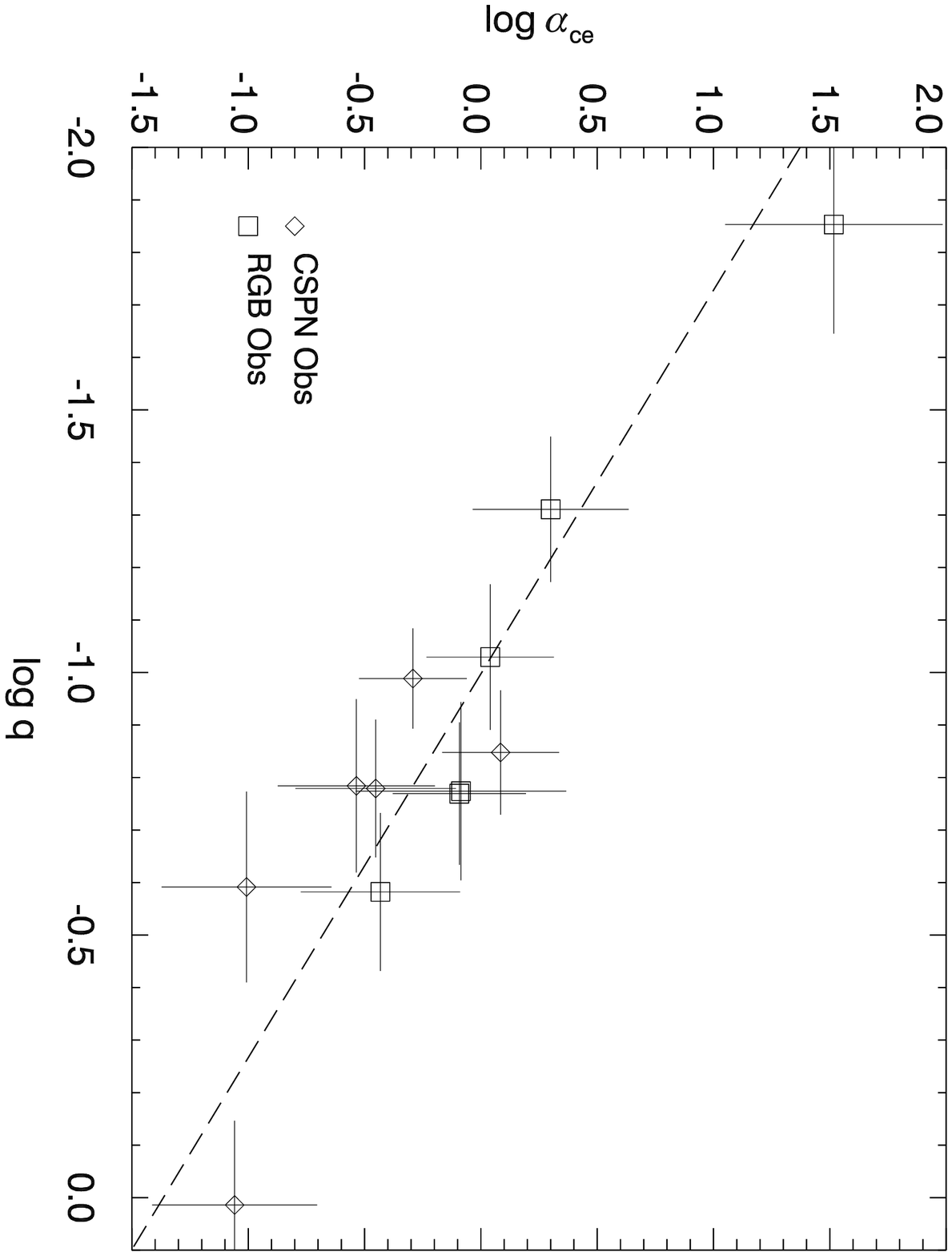}
 	\end{center}\caption{A plot  of $\log (q = M_2 / M_1$) vs. $\log$ \ace. The simulations (crosses) are not fitted as the errors could not be determined. Top panel: all data is presented. Bottom panel: same as the top panel but without the post-AGB systems with no PN. The dashed lines are least square fits of the data.}
	\label{fig:alpha}
\end{figure*}

A fit to $\log$\ace\ as a function of $\log (q=M_2/M_1)$ is plotted in  Fig.~\ref{fig:alpha}.  In Table~\ref{tab:statistics} we list the parameters of the linear fit with their errors, along with the number of data points, reduced $\chi^2$ value, probability to exceed\footnote{The probability to exceed is the probability that a set of randomly sampled data from a normal distribution produces a reduced $\chi^2$ that exceeds the one determined. This value should be as close as possible to 0.5, corresponding to a reduced $\chi^2$ of approximately unity, where the exact value depends on the number of degrees of freedom \citep{Bevington2003}.} and the correlation coefficient $r$; we then list constant \ace\ fits, along with their reduced $\chi^2$ and probability to exceed. First, we fit all of our data together, except for the simulations, as before. The data is reasonably anti-correlated. We then exclude the outlier (HD~149382 [$q=0.014$, \ace = 33]); the anti-correlation is somewhat degraded, although these two fits are very equivalent from a statistical point of view. Both have a reduced $\chi^2$ below unity and a probability to exceed of 0.8, meaning that the errors may have been overestimated or too many degrees of freedom have been considered in the fit. However, a constant \ace\ fit never represents the data better with a probability to exceed of zero, which means that random deviations would always explain the data better than the constant \ace\ fit. 

%We also fit the data excluding the two outlier data points (AA~Dor [$q= 0.006$, \ace = 128] and HD~149382 [$q=0.014$, \ace = 33]).

To investigate further our data, we eliminated groups of systems to determine whether this would improve the statistics of the anti-correlation among the remaining data. By fitting only post-RGB and central stars of PN systems can one improve the anti-correlation (Fig.~\ref{fig:alpha} -- right panel).  On the other hand the probability to exceed reduces further and since the constant \ace\ fit is not better, we once again have to conclude that the errors have been overestimated for this fit. The improved anti-correlation might derive from the fact that in our reconstruction technique we have assumed that any primary with mass larger than 0.47~\msun, went through the AGB evolution. That assumption results in a specific primary mass at the time of the CE interaction (\S~\ref{sssec:mass}).  It is however possible, that some of these primaries are actually post-RGB objects descending from {\it more massive} stars, as we described in \S~\ref{ssec:reconstruction}. More massive stars develop a helium core more massive than 0.47~\msun\ before ascending the RGB. Once they expand, they may suffer a CE interaction on the RGB  that truncates the RGB evolution by ejecting the envelope. These massive, post-RGB, post-CE primaries would later ignite core helium, thus developing a CO core, but with an envelope mass low enough to make them appear as hot sub-dwarfs with a close companion. It is therefore possible that these objects' masses and radii at the time of the CE interaction might have been significantly miscalculated, resulting in the observed large scatter on the $\log q$ vs. $\log$ \ace\ plot in Fig.~\ref{fig:alpha} (left panel). On the other hand, the central stars of PN are (almost) guaranteed to have been on the AGB, and the systems with $M_c < 0.47$~\msun\ are post-RGB stars, making these two groups less prone to error.

%We have also calculated the values of \ace\ following the reconstruction techniques of \citet[][]{Afsar2008},  for  systems with primary masses outside the range $\sim$0.47 - 0.55~\msun\ -- masses within that range cannot be used with their treatment because they would imply either a negative luminosity or have enormous errors, as explained in \S~\ref{ssec:reconstruction}. The values of \ace\ are only marginally larger than in our treatment (20-50\%), but the anti-correlation trend is not present. A closer inspection reveals that while their values of the primaries' radii at the time of the CE interaction are similar to our own (within the error), the values of the primaries' masses determined with their treatment are too large (around 3~\msun) and are inconsistent with the IMF and the IFMR (they used the $M_c$ vs. $M_1$ relation of \citet{Iben1986}). In addition, their values of $\lambda$, calculated with the formula of \citet{Webbink2008}, are all larger than unity, in conflict with the values determined by our extensive analysis. The larger values of $M_1$ and $\lambda$ in their treatment conspire to maintain the values of \ace\ within a range similar to our own. We believe that our own treatment, where every step is rooted in a rigorous treatment of stellar structure, together with our extensive error analysis, leads to a more trustworthy calculation of the values of \ace. 

The values of \ace\ derived by \citet{Sandquist1998} are approximately a factor of four lower than those we derive. They used the formalism of \citet[][]{Tutukov1979a}, except that they revised the expression for the orbital energy to include the orbital energy value at the time of the CE interaction, unlike \citet{Tutukov1979a}. \citet{Sandquist1998} also derived the binding energy as the {\it difference} between the binding energy at the beginning and at the end of the CE interaction, where they integrated the stellar structure for the initial value and used the simulation output to determine the remaining binding energy from the left over bound envelope in the system. Although we suspect that the leftover envelope binding energy is low, they do not report the actual value. It is not possible at this point to account exactly for this discrepancy. However a large part of it is certainly due to  the fact that \citet{Tutukov1979a}, like \citet{Yungelson1993} (Eqs. \ref{eq:tutu} and \ref{eq:Yun}) use a far larger radius of the primary at the time of the CE interaction, reducing the value of \ace\ (see \S~\ref{ssec:literature}).

% =============================================================================
%							THERMAL ENERGY
% =============================================================================

\section{The stellar response and the thermal energy}
\label{sec:discussion}

In \S~\ref{ssec:virial} we discussed using the virial theorem to quantify the entire energy budget of the envelope. If we accounted for the thermal energy of the envelope in this way, we would multiply the envelope binding energy by a factor of 0.5, which would lower all $\alpha$ values by a factor of two, and bring all but one of the observed systems below unity. 
%HD149382 remain enigmatic. The fact that \ace\ for these two systems is much larger than unity, even when we account for the thermal energy of the gas via the virial theorem, may point to other energy sources. Perhaps dissociation and ionisation energy of the molecules and atoms, as already discussed by \citet{Han1995} and \citet{Webbink2008}, can be more efficiently tapped by smaller companions. 

It is possible that the anti-correlation in the data, where some of the systems with low $q$ value appear to {\it need} the thermal energy, while others do not, may hold some clues. Since the virial theorem is based on the principle of hydrostatic equilibrium, and since adjustments to hydrostatic equilibrium take place on the stellar dynamical timescale, we may hypothesise that the timescales of envelope penetration and in-spiral vary as a function of $q$.

The equation of hydrostatic equilibrium dictates that, during their giant phases, stars expand upon losing mass because the pressure gradient overcomes gravity. When the envelope mass is reduced below a certain threshold by mass-loss, the gravity term dominates and the star shrinks rapidly - this is the phase when the star leaves the RGB or AGB and moves to the left of the Hertzsprung-Russell diagram. If a small companion triggers additional mass-loss while in the outer parts of the envelope, it will also trigger expansion and reduction of the envelope binding energy, because of radius relaxation. This means that relatively light companions could unbind an envelope that is in principle too heavy for them to lift.  

The time over which stars respond to changes is the stellar dynamical time. For a 100-\rsun\ RGB star this is approximately two weeks, while for a 500-\rsun\ AGB star it is of the order of a few months to a year. The latter is similar to the CE interaction timescale determined by the few simulations carried out so far \citep[e.g.,][]{Sandquist1998,DeMarco2003}.
On these grounds {\it we suggest that if the in-spiral of the companion takes longer than a dynamical time the giant can use its own thermal energy to help unbind the envelope, but if it takes less, this energy source will not be available}. It may be possible that when $q$ is smaller it takes the companion longer to plunge into the primary envelope, thus giving the primary star more time to respond by expanding and reducing its own envelope binding energy. In other words, for slower changes the star has time to become virialised, which results in a stellar expansion, i.e., the star is tapping its own thermal energy.

This idea is supported by the calculation of \citet[][their Figure 2]{Nordhaus2006} that shows that lower mass companions take longer to in-spiral. This also makes sense in view of the fact that the transfer of orbital energy to the envelope is via gravitational drag \citep{Ricker2008}: the companion creates a dense wake that slows it down and makes it fall in towards the giant's core. Such a wake would be less massive for lower mass companions. This suggestion can also explain the simulations' above-unity \ace\ values. Hydrodynamic simulations do not include the stellar energy source, but as long as the CE in-spiral timescale is shorter than the thermal timescale, the simulated star will behave physically.

HD~149382, with a sub-stellar companion, remains troublesome. Its value of \ace\ is so high that it cannot be explained simply by invoking the thermal response of the star. One could invoke additional sources of energy such as dissociation and ionisation energies, but it seems suspicious that such source should only be invoked for the lightest companion. It is possible that a third body was present in these systems, such as another planet. We could also question the precision of the determined parameters; however we note that with such a low mass companion, there is no reasonable way of reducing the value of \ace. To test this we increased that primary mass from 0.41~\msun (the middle of the calculated range) to 0.47~\msun,  suggested by \citet{Geier2009} as a more likely value and we determined \ace\ under the assumption that HD~149382 is a post-RGB system as well as a post-AGB one. The smallest value of \ace\ (10) is obtained for a post-AGB object and the reduction from the value of 33 (Table~\ref{tab:alphaoutputs}) is due primarily to the fact that the radius would be larger for a larger core mass. In our log--log fit this does not alter appreciably the fit parameters and the anti-correlation is preserved.

 % =============================================================================
%							CONCLUSION
% =============================================================================

\section{Summary and discussion}
\label{sec:summary}

In this paper we have compared different derivations of the CE efficiency equations used in the literature, and have derived a more accurate form. Our approximation is quite similar to that of \citet{Webbink1984}, although our binding energy is more negative, resulting in higher values of \ace. 

We have then used this approximation to determine the value of \ace\ for a set of observed post-CE systems as well as a set of simulations. In so doing we have revealed the difficulties inherent in deriving \ace\ values from observations:  the methods available to reconstruct the parameters of the primary at the time of the CE interaction do not lead to accurate values. As a result the errors on the determined \ace\ are large. 

We have also found some evidence for an anti-correlation between $\log (q = M_2/M_1)$ and $\log$~\ace, such that $\alpha$ $\approx$(0.05 $\pm$ 0.02)$\times q^{(-1.2 \pm 0.4)}$.  Considering the exponent of $q$ is $\la -1$, this implies that smaller mass companions are left with comparable, if not longer post-CE periods than their more massive counterparts. One way this could be achieved is if smaller companions took longer than the primary's dynamical time to penetrate into the CE than larger ones, giving their primary star an opportunity to react and use its own thermal energy to expand, thus helping to eject the envelope. Since the CE timescales seem to be similar to the stellar dynamical time scale \citep[of the order of a year; ][]{Sandquist1998,DeMarco2003} this suggestion is plausible. 

The energetics of the CE interaction are complex, however, and it is not enough to state that ionisation and dissociation energies could be available, one has to also show that once tapped, these sources aid the envelope ejection process, a thing that was questioned by \citet{Soker2003} who argued that the opacity of the envelope would be too low. Several other mechanisms could play a role in the common envelope ejections \citep[e.g., excitation of pulsational waves][]{Soker1992}, which may also depend on the in-spiral  time scales.

\citet{Politano2007} and \citet{Davis2010} calculated population synthesis models under the assumption of a constant \ace, or \ace\ proportional to secondary mass. \citet{Politano2007} compared the predicted post-CE period distribution, secondary and primary mass distributions with observations of present day post-CE systems and cataclysmic variables. The predicted distributions are not strongly dependent on the prescription of  \ace\ they used, leaving \ace\ unconstrained. Their period distribution for current day post-CE systems predicts many more systems at longer periods than observed (e.g., Miszalski et al. (2009) or \citet{Schreiber2008} and \citet{Schreiber2009}). The predicted secondary mass distribution seems also in contrast with observations that show that the most represented companion mass has spectral type M3.5, similarly to the distribution of field main sequence stars \citep{Farihi2005}. \citet{Politano2007} claim that the large fraction of brighter companions they predict would be missed by surveys because they would outshine the WD primary. However \citet{Holberg2008} shows that the companions that would be missed because they outshine the primary WD are brighter than mid-to-late K. This means that if the real companion mass distribution were that predicted by the models of \citet{Politano2007}, we would have detected more companions with mass larger than $\sim$0.4~\msun. 

\citet{Davis2010} predict the space density of post-CE systems, the distribution of secondary masses and the period distribution. The latter two predictions do not match observations. \citet{Davis2010} claim that similarly to observations they predict a steep decline of the systems with secondary masses $<$0.35~\msun. However, as we can see from the exhaustive compilation of \citet{Zorotovic2010}, the number of systems with secondary mass below that threshold is very large. Finally, as is the case for the \citet{Politano2007} models, the predicted period distribution peaks at longer periods and is very broad, contrary to the observed ones, whose distribution peaks at periods shorter than a day and is very narrow. The period distribution is likely to become the best observation against which to test our population synthesis models in the future.

\subsection{A comparison of this work to that carried out by Zorotovic et al. (2010)}
\label{ssec:Zorotovic}

While this paper was in review we became aware of a publication by Zorotovic, Gaensicke \& Schreiber (2010), in which the authors determined the values of \ace\ for a set of post-CE WDs and drew conclusions from their findings in much the same way as we have done here. Some of their conclusions are similar to our own, while others differ.

Both our and their data derive from that of \citet{Schreiber2003}. However, they added 35 systems that their group has newly discovered. On the other hand, they excluded subdwarf primaries, such as all central stars of PN, and the systems with low mass secondaries (AA~Dor and HD~149382). We have insured that for all the objects in common we used the same parameters.

\ZT\ remarked that \ace\ values from simulations are in the range determined by their technique for the observed systems. We maintain instead that there is no way of comparing \ace\ values from simulations to those obtained by system reconstruction, without fully understanding what \ace\ formalism was adopted by the simulation studies and subject to the caveats of the simulations. This is why we have selected those simulations that we understand the best and re-determined their \ace\ values using the same formalism used for the observed systems.

\ZT\ compared three \ace\ formalisms, the \citet{Yungelson1993} formalism, the \citet{Webbink1984} formalism and a hybrid form between the two formalisms adopted by \citet{Hurley2002}. Differences between \ace\ values obtained with these formalism are in line with our discussion in \S~\ref{ssec:literature}. However, in contrast to \ZT\ we have argued that it is possible to select the best formalism on physical grounds. In the end they adopted the formalism of  \citet{Webbink1984}, making a comparison with our own work straight forward.

We agree with \ZT\ on the treatment of $\lambda$, i.e., that one should calculate it according to what stellar structure one is examining. On the other hand, they did not disclose what values they used for those cases when they calculated $\lambda$ using stellar structure calculations, only mentioning that the calculation was performed. They used the hydrogen abundance criterion for the core-envelope boundary, rather than the remnant envelope criterion, but that does not change significantly the resulting value of $\lambda$.

The primary reconstruction technique is usually the hardest part to compare. \ZT\ used the stellar evolution fitting formulae of \citet{Hurley2000} to determine core mass, radii and luminosities as a function of time for a range of main sequence masses. Each measured primary mass ($M_c$) has multiple matching parameter sets ($L$, $R$ and $M_{\rm MS}$); each set can be used to determine \ace. In this way each observed system will have a range of \ace\ values. We, on the other hand, determined only one value of \ace, but our error bar encompasses similar ranges to those determined by \ZT. 

For example, \ZT\ did not use the IFMR to determine $M_{\rm MS}$ for their systems, but instead determined a range of main sequence masses which, sometime during their evolution, will attain the same core mass as that of the primary of the observed systems. This is equivalent to using the IFMR implied by the fitting formulae, along with the uncertainty in those very relations and the measurement errors (the error propagation is not discussed in their paper, nor the uncertainties in the fitting relations). In the end, the largest  difference between the two methods is likely to lie in the difference between the \citet{Hurley2000} fitting formulae and our MESA evolutionary sequences and fits to the \citet{Bertelli2008} tracks.

The \ZT\ reconstruction technique does not pre-select whether the system went through a CE interaction on the RGB or AGB. Indeed several of their systems may have gone through such interaction in either phase.  We find that all our post-RGB systems are post-RGB systems also in their scheme, while  those of our systems with $M_c$ in the range 0.47-0.55~\msun, which are post-AGB in our scheme (recall our criterion: $M_c >0.47$~\msun), have a chance to have gone through a CE interaction on the RGB. We pointed out (\S~\ref{sssec:anti-correlation}) that our conclusions remain true, and in fact become even stronger if we eliminate the entire class of post-AGB stars with no PN, leaving the central stars of PN (not addressed by \ZT) and the post-RGB stars (which are post-RGB stars also in \ZT).

After determining ranges of \ace\ values for each star, \ZT\ determined the most likely values within those ranges by carrying out a population synthesis exercise. We have effectively carried out a similar exercise by using Baysian statistics in the determination of mass and radius of the primary at the time of the CE, by  calculating the differences between $M_{\rm MS}$ and $M_1$ and between $M_{WD}$ and $M_c$ using statistical arguments. Interestingly, the conclusions drawn in the two papers are similar: \ace\ values for post-RGB stars are higher. Avery weak to no correlation of \ace\ with $M_2$ exists (Fig.~\ref{fig:alpha2}, middle panel). Post-AGB systems display a mild correlation between period and \ace, while post-RGB systems do not (Fig.~\ref{fig:alpha2}, bottom panel). \ZT\ did not test possible correlation between $q$ and \ace\ nor did they carry out a statistical treatment of their data. In addition, they did not list their mean \ace\ values nor the corresponding stellar parameters, making it difficult to carry out a more thorough comparison.

%---------------
%APPENDIX A
%---------------

\appendix
\section{The core-envelope boundary and the value of $\lambda$ for different stellar models and evolutionary stages.}\label{appendixA}

\begin{figure*}
\vspace{10cm}
	\begin{center}		
		\includegraphics{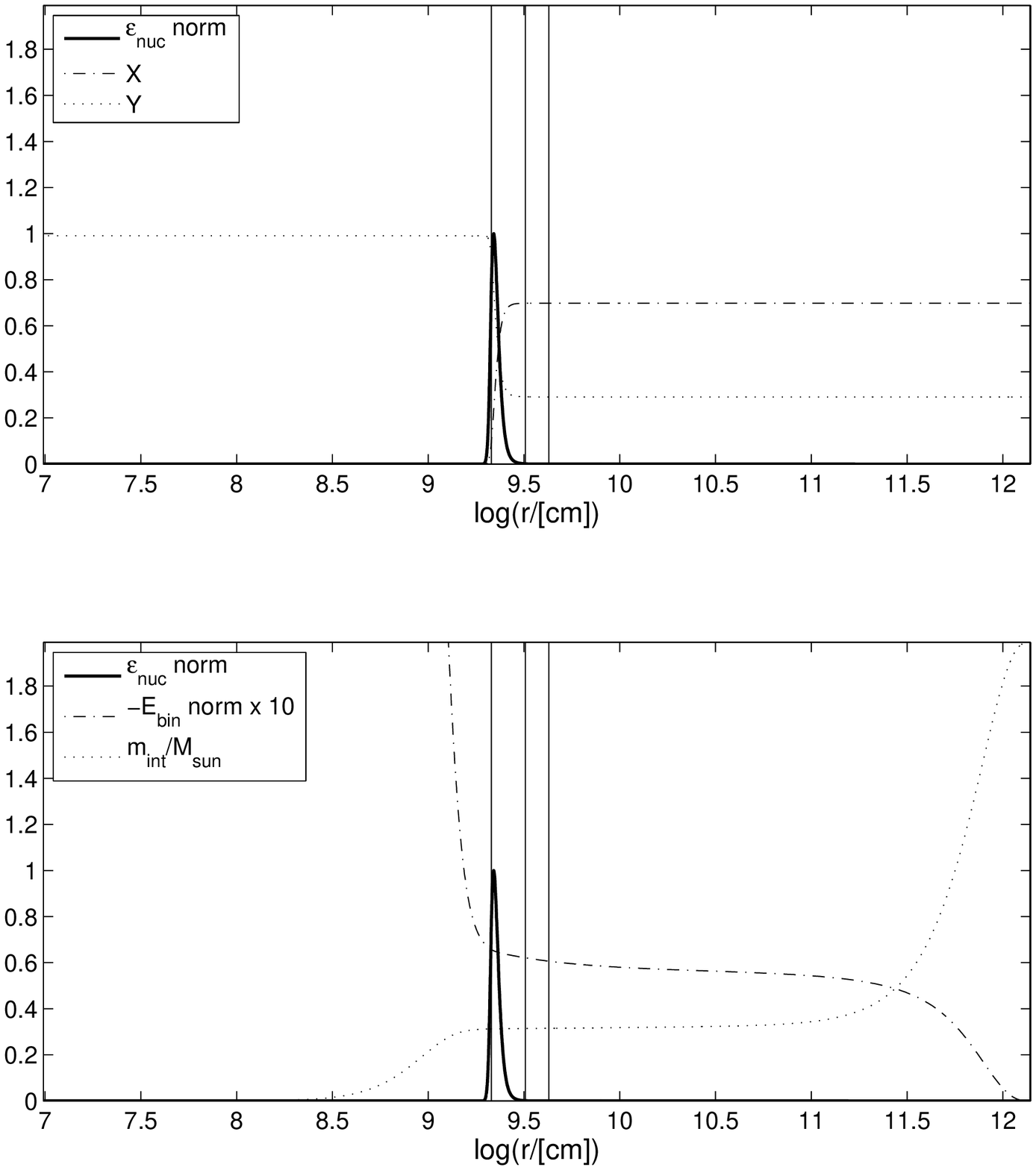}
                   \includegraphics{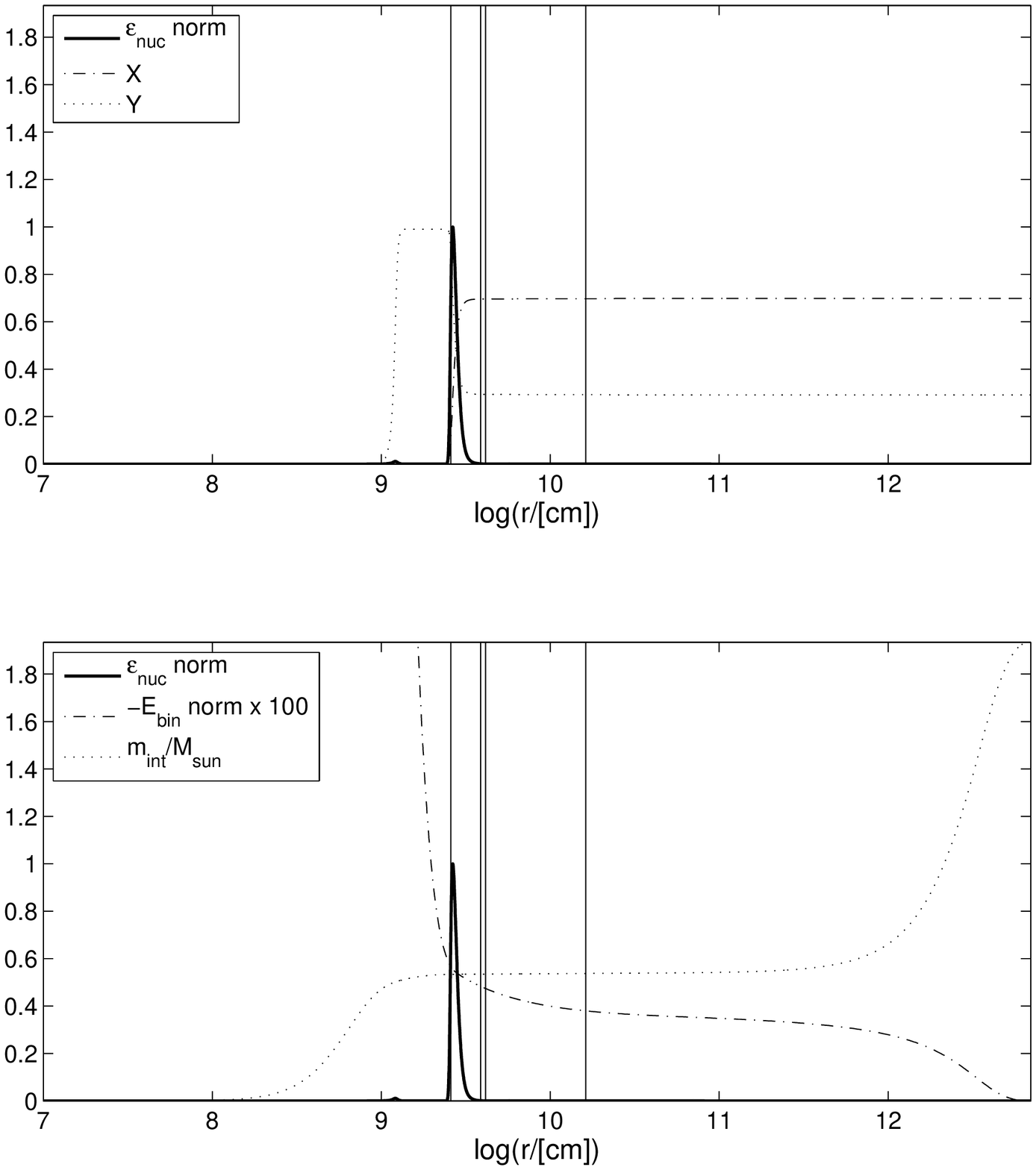}
	\end{center}\caption{Stellar structures for an RGB star (left panels) which had a main sequence mass of 2\msun\ and its AGB counterpart (right panels; for additional model parameters see Table~\ref{tab:lambda}). Vertical solid lines represent the core/envelope boundaries chosen according to the ``$X_H=0.1$" criterion (leftmost vertical line), ``shell" criterion and ``remnant envelope" criterion (second and third vertical lines from the left - overlapping in the left panels) and ``density" criterion (rightmost vertical line). }
	\label{fig:CoreBoundary}
\end{figure*}
To locate the core-envelope boundary one could use a criterion based on density alone (called ``Density" in Table~\ref{tab:lambda}), such as the radius value that minimises ${dm}/{d\log r }$. This criterion leads to a larger radius than any criteria based on a physical argument. Aside from the criterion we have adopted (dubbed ``Remnant Envelope" in Table~\ref{tab:lambda}, see also \S~\ref{ssec:lambda}), there are two additional criteria that lead to similar values of $\lambda$. (i) The radius where the abundance of hydrogen is 0.1 (this is the criterion adopted by \citet{Dewi2000}), and (ii) the largest radius where the nuclear reaction rate decreases below a given threshold value  (just outside the hydrogen-burning shell; dubbed ``Shell" in Table~\ref{tab:lambda}). 

\begin{table}
\begin{tabular}{lcccc}
%\tabletypesize{\scriptsize}
\hline
Criterion:&\multicolumn{1}{c}{Density}&\multicolumn{1}{c}{$X_H$=0.1}&\multicolumn{1}{c}{Shell  }&\multicolumn{1}{c}{Remnant Env.}\\
\hline
&\multicolumn{4}{c}{RGB}\\
%$M_c$(\msun)  &  0.308  &  0.283  &  0.291  &  0.292    \\
%$R_c$(\rsun)  &  0.238  &  0.0436  &  0.0669  &  0.0729    \\
%$X_H$  &  0.30  &  {\bf 0.1}  &  0.26  &  0.26    \\
%$E_{bin}/10^{47}$(cgs)  &  $-$8.99  &  $-$12.03  &  $-$10.63  &  $-$10.43    \\
%$\lambda$  &  0.49  &  0.37  &  0.42  &  0.43    \\
%Ind  &  1058  &  1335  &  1315  &  1133    \\
$M_c$(\msun)  &  0.314  &  0.313  &  0.313  &  0.314    \\
$R_c$(\rsun)  &  0.044  &  0.031  &  0.032  &  0.040    \\
$X_H$  &  0.70  &  {\bf 0.1}  &  0.23  &  0.69    \\
$E_{\rm bin}/10^{47}$(ergs)  &  $-$9.14  &  $-$9.62  &  $-$9.55  &  $-$9.23    \\
$\lambda$  &  0.40  &  0.38  &  0.38  &  0.39    \\
\hline
&\multicolumn{4}{c}{AGB}\\
%$M_c$(\msun)  &  0.499  &  0.494  &  0.496  &  0.496    \\
%$R_c $(\rsun) &  0.253  &  0.0412  &  0.0744  &  0.0804    \\
%$X_H$  &  0.70  &  {\bf 0.1}  &  0.70  &  0.70    \\
%$E_{bin}/10^{47}$(cgs)  &  $-$2.47  &  $-$3.49  &  $-$2.88  &  $-$2.83    \\
%$\lambda$  &  0.41  &  0.29  &  0.35  &  0.35    \\
%Ind  &  653  &  1322  &  1306  &  1289    \\
$M_c$(\msun)  &  0.537  &  0.534  &  0.534  &  0.534    \\
$R_c $(\rsun)  &  0.233  &  0.037  &  0.038  &  0.040    \\
$X_H$  &  0.70  &  {\bf 0.1}  &  0.28  &  0.52    \\
$E_{\rm bin}/10^{47}$(ergs)  &  $-$2.05  &  $-$3.06   &  $-$3.00   &  $-$2.92     \\
$\lambda$  &  0.32 &  0.21  &  0.22  &  0.22    \\
\hline
\end{tabular}
\caption{Parameters at the core-envelope boundary according to different criteria for the 2-\msun\ model on the RGB ($M_1$=2.0~\msun, $R = 20$~\rsun) and AGB ($M_1$=1.93~\msun, $R = 100$~\rsun).\label{tab:lambda}}
%\tablenotetext{a}{}
\end{table}

Although the abundance criterion leads to a lower value of the core-envelope boundary radius, a glance at Table~\ref{tab:lambda}  shows us that the resulting values of $\lambda$ are quite similar.
There we list the values of the core mass, core radius and hydrogen abundance of the 2-\msun\ MESA model for two key evolutionary phases, when the star is half way up the RGB (M=2.0~\msun, R=20~\rsun\ and $\tau$=1.0~Gyr) and at the start of the thermally-pulsating AGB (M=1.93~\msun, R=100~\rsun\ and $\tau$=1.2 Gyr), along with the derived values of the envelope binding energy and of $\lambda$. 

\section*{Acknowledgments} OD and M-MML acknowledge funding from NSF grant AST-0607111. Eric Sandquist is thanked for useful discussions on their stellar structure models. OD is grateful to Lev Yungelson for clarifying the history of the \ace\ equation. Gijs Nelemans is thanked for useful comments. The referee, Zhanwen Han is thanked for in depth comments that helped improve this paper. The stellar evolutionary track part of the work was also supported by the National Science Foundation under grants PHY 05-51164 and AST 07-07633. FH acknowledges funding through a NSERV Discovery Grant.

\bibliography{convert}
%\bibliography{../../../../../convert,../../../../../bibliography}

\label{lastpage}
\end{document}